\documentclass[fleqn,usenatbib]{mnras}

\usepackage{newtxtext,newtxmath,mathtools,amssymb}

\usepackage[T1]{fontenc}
\usepackage{multirow}
\usepackage{graphicx}	
\usepackage{amsmath}	
\usepackage{amssymb}	
\usepackage{ae,aecompl}
\usepackage{euscript}
\usepackage{setspace}
\usepackage{bigints}
\usepackage{booktabs,makecell}
\usepackage{caption}
\usepackage{subcaption}
\usepackage{multirow}
\usepackage{soul}
\usepackage{tikz}
\newcommand{\myarrow}{\tikz\draw[thin,black,-latex] (0,2.5ex) -- ++(0,-1.5ex) -- +(2.5ex,0);}

\title[UHE Particles in Protogalactic Environments]{Interactions between Ultra-High-Energy Particles and Protogalactic Environments}

\author[Owen et al.]{
Ellis R. Owen,$^{1,\;\! 4}$\thanks{E-mails: ellis.owen.12@ucl.ac.uk (ERO), idunn.jacobsen.09@ucl.ac.uk (IBJ), kinwah.wu@ucl.ac.uk (KW), 
  pooja.surajbali@mpi-hd.mpg.de (PS)}
Idunn B. Jacobsen,$^{1}$
Kinwah Wu,$^{1}$
Pooja Surajbali\hspace{2pt}$^{2,\;\! 3}$
\\
$^{1}$Mullard Space Science Laboratory, University College London, Holmbury St. Mary, Dorking, Surrey, RH5 6NT, United Kingdom\\
$^{2}$Department of Physics \& Astronomy, University College London, Gower Street, London, WC1E 6BT, United Kingdom\\
$^{3}$Max-Planck-Institut f\"{u}r Kernphysik, Saupfercheckweg 1, Heidelberg 69117, Germany\\
$^{4}$Institute of Astronomy, Department of Physics, National Tsing Hua University, Hsinchu, Taiwan (ROC)
}

\date{Accepted XXX. Received YYY; in original form ZZZ}

\pubyear{2018}

\begin{document}
\label{firstpage}
\pagerange{\pageref{firstpage}--\pageref{lastpage}}
\maketitle
\begin{abstract}
We investigate the interactions of energetic hadronic particles (cosmic ray protons) 
  with photons and baryons in protogalactic environments, where  
the target photons are supplied by the first generations of stars to form in the galaxy and the cosmological microwave background, 
  while the target baryons are the interstellar and circumgalactic medium. 
We show that pair-production and photo-pion processes are the dominant interactions 
  at particle energies above $10^{19}\;\! {\rm eV}$, 
  while ${\rm pp}$-interaction pion-production dominates at the lower energies 
  in line with expectations from, 
  for example $\gamma$-ray observations of star-forming galaxies and dense regions of our own galaxy's interstellar medium.  
We calculate the path lengths for the interaction channels 
   and determine the corresponding rates of energy deposition. 
We have found that protogalactic magnetic fields and their evolution 
   can significantly affect the energy transport and energy deposition processes of cosmic rays. 
Within a Myr after the onset of star-formation  
  the magnetic field in a protogalaxy could attain a strength  
  sufficient to confine all but the highest energy particles within the galaxy.  
This enhances the cosmic ray driven self-heating of the protogalaxy 
   to a rate of around $10^{-24}\;\! {\rm erg} \;\!{\rm cm}^{-3}\;\! {\rm s}^{-1}$ 
   for a galaxy with strong star-forming activity that yields 1 core collapse SN event per year. 
This heating power exceeds even that due to radiative emission from the protogalaxy's stellar populations. 
However, in a short window before the protogalaxy is fully magnetised,   
  energetic particles could stream across the galaxy freely, delivering energy into the circumgalactic and intergalactic medium. 
\end{abstract}




\section{Introduction}
\label{sec:introduction}

High-energy hadronic particles heat the media along their propagation paths  
  by collisional ionisation and hadronic interactions, 
  with such action of cosmic rays (CRs) having been observed close-by in the Earth's atmosphere
  \citep[see e.g.][for reviews]{Ginzburg1996PhyU, Kotera2011ARA&A}. 
A substantial fraction of the energetic CRs detected on Earth are protons \citep{Abbasi2010PRL}, 
  and their origins are likely to be extragalactic.  
The interactions of CRs in astrophysical environments have also been discussed in the literature in various contexts 
   \citep[e.g.][]{Greisen1966PhRvL, Zatsepin1966JETPL, McCray1972ApJ, Nath1993MNRAS, Valdes2010MNRAS}. 
While research into CRs in solar-terrestrial settings has progressed a long way since their discovery,   
  the role of ultra-high-energy (UHE) CR particles in galaxies and larger structures 
   is less well understood --  
   even in systems where these energetic particles are evidently abundant, such as in star-forming galaxies 
   \citep[see e.g.][]{Karlsson2008, Lacki2011ApJ, Lacki2012AIP, Wang2014} 
   and galaxy clusters \citep[see e.g.][]{Brunetti2014}.     
It has been argued that CRs can regulate star-formation in a galaxy \citep[see][]{Pfrommer2007, Chen2016} 
  and that their heating effect could drive large-scale galactic wind outflows \citep[see][]{Socrates2008, Weiner2009ApJ}.    
Not only do UHECRs alter the dynamical and thermal properties of their galaxy of origin, they can also transport energy from their host to the surrounding intergalactic medium.  
While CR heating and ionisation operate in nearby galaxies, conditions in the very distant Universe imply that CR processes are expected to be even more important 
   when young galaxies were spawning their first generation of stars 
   and luminous quasars first emerged.      
The effects of CRs in the high-redshift Universe are now beginning to be recognised  
   \citep[see][]{Giammanco2005A&A, Stecker2006ApJ, Valdes2010MNRAS,  
   Bartos2015, Sazonov2015MNRAS, Walker2015arXiv, Leite2017MNRAS}.  
 
CRs are products of violent astrophysical events,  
  e.g.\ supernova (SN) explosions, gamma-ray bursts, 
  large-scale shocks in colliding galaxies and galaxy clusters,  
  and extreme environments in compact objects, 
  e.g.\ fast spinning neutron stars and accreting black holes 
  \citep[see][]{Berezinsky2006, Dar2008PhR, Kotera2011ARA&A}. 
As strong star-formation activity gives rise to frequent SN events,   
   galaxies with active, ongoing star-formation are naturally strong CR sources. 
In these galaxies, the stars and their remnants supply the seed particles, 
   and the shocks generated by the SN explosions and other violent events provide the needed mechanisms 
   for accelerating the particles to very high energies, e.g. through Fermi acceleration processes \citep{Fermi1949PR}. 

Atomic matter can be ionised and excited by keV CR protons and this can heat the interstellar and intergalactic medium  
 \citep[see e.g.][]{Nath1993MNRAS, Sazonov2015MNRAS}.  
However, compared to their higher-energy counterparts, 
    a galactic population of sub-GeV CRs only contains sufficient energy to drive a relatively small heating effect, 
    even if all of their energy were to be deposited within a reasonable timescale. 
This is because less than 1\%  of the total CR energy is harboured in these lower energy particles 
  \citep[see, e.g.][]{Benhanbiles-Mezhoud2013}. 
Higher-energy CRs (i.e. those with energy $\gtrsim 1 \;\!{\rm GeV}$) are much less engaged 
  with atomic interactions.  
If ionisation and atomic excitation were the only means of energy exchange, 
  a 0.5 GeV CR-proton would lose less than $2.5$\% of its initial energy 
  on a Hubble timescale 
  as it propagates across a cosmological baryonic density field,  
  even when the density is maintained as that at redshift $z = 20$ \citep[see][]{Sazonov2015MNRAS}. 
However, 
  the distance a CR particle above GeV energies can propagate in an interstellar or intergalactic medium 
  is instead determined by hadronic processes.     
The Gresisen-Zatsepin-Kuzmin cutoff  \citep{Greisen1966PhRvL, Zatsepin1966JETPL} 
  is an example of such, 
  where CRs are suppressed by their hadronic interactions 
  with cosmic microwave background (CMB) photons.  

The impact of CRs depends on their interaction channels with the matter and radiation fields 
  (microscopic particle physics) 
  and also on the properties of the media through which the CRs propagate (microscopic astrophysics).    
  How CRs affect the interior and exterior environments of their host galaxy  
  is determined by the ability of the galaxy to act as a CR calorimeter (accounting for the galactic density profile and substructure),
  its regulation of CR diffusion, 
  the properties of its global outflows 
  and the entrapment of charged particles by its magnetic field 
  \citep[see][]{Thompson2007, Tueros2014A&A, Kobzar2016ArXiv}.

In this work we investigate the effects of energetic CR-particles in high redshift protogalactic environments.  
We adopt a phenomenological approach that sufficiently captures 
  the essence of the relevant physics and astrophysics.  
Complexities such as the CR compositions at different stages of galactic evolution 
  or the exact production and acceleration mechanisms of various types of primary particles 
  in specific galactic components 
  are finer details that will be left to future follow-up studies.  
Our focus is on the interactions of the particles with the radiation and baryon density fields 
 and how efficiently energy is deposited 
 along the particle propagations in the interstellar and intergalactic media.  

We organise the paper as follows. 
In \S~\ref{sec:interactions} 
 we introduce the hadronic processes relevant to CR heating of the astrophysical media. 
In \S~\ref{sec:protogalaxies}, 
 we specify the protogalactic model that provides the baryonic density profile, radiation properties and magnetic fields  
 through which the propagations and interactions of the CRs are calculated. 
In \S~\ref{sec:discussion}, we show the results of our calculations, and we discuss their consequences and astrophysical implications in \S~\ref{sec:conclusions}. We particularly focus on those implications concerning propagation distances and energy deposition rates of the CRs 
 and particle confinement in response to the evolution of the magnetic field of the host galaxy.

For the remainder of this discussion, we refer to the UHECR protons as CRs, unless stated otherwise. 
For clarity we may also refer specifically to CR protons and CR electrons to differentiate between the two where appropriate, 
  but when unspecified we refer to the proton component of cosmic rays. 
We do not consider primary CR electrons in detail as their energy loss rate is considerably more rapid compared to the CR protons. 
Therefore, their energy is deposited much more locally to their source and is less important in the global model we present here.

\section{Particle Interactions}
\label{sec:interactions}

Photo-pion (${\rm p}\gamma$) production, Bethe-Heitler pair production and ${\rm pp}$ pion-production interactions 
  are the relevant processes 
  governing the interaction of hadronic CRs with radiation and matter in astrophysical environments.  
These processes produce a variety of hadrons, including charged and neutral pions, neutrons and protons   
    \citep[see][]{Pollack1963PhRv, Gould1965AnAp, Stecker1968ApJ, 
     Mucke1999PASA, Berezinsky2006, Dermer2009book}.      
The secondary particles produced will continue to undergo hadronic interactions,   
   resulting in particle cascades.     
Neutral pions predominantly decay into two photons through an electromagnetic process,  
\begin{align}%
	\pi^0	&\rightarrow 2\gamma  \ ,   %
\end{align}%
  with a branching ratio of 98.8\%~\citep{Patrignani2016ChPh} and on a timescale of $8.5 \times 10^{-17}\;\!{\rm s}$.   
Charged pions predominantly produce leptons and neutrinos via a weak interaction,   
 \begin{align}%
 \label{eq:weak_interaction2}
	\pi^+	&\rightarrow \upmu^+ \nu_{\rm \upmu} \rightarrow \rm{e}^+ \nu_{\rm e} \bar{\nu}_{\rm \upmu} \nu_{\rm \upmu}\     \nonumber \\%
	\pi^-	&\rightarrow \upmu^- \bar{\nu}_{\rm \upmu} \rightarrow \rm{e}^- \bar{\nu}_{\rm e} \nu_{\rm \upmu} \bar{\nu}_{\rm \upmu}       %
 \ , 
\end{align}%
 with a branching ratio of 99.9\%~\citep{Patrignani2016ChPh} but on a longer timescale of roughly $2.6\times 10^{-8}\;\!{\rm s}$.  

As the majority of the CR-particles observed on Earth are protons \citep[see][]{Abbasi2010PRL},   
   we boldly assume that the particles and the baryons in the protogalactic environment are protons. 
The energy of the incident particles is $E_{\rm p} = \gamma_{\rm p} m_{\rm p} \rm{c}^2$, 
   where $\gamma_{\rm p}$ is the Lorentz factor.  
When normalised to the electron mass this   
  is $\epsilon_{\rm p} = \gamma_{\rm p} (m_{\rm p}/m_{\rm e})$.    
The target baryons may be assumed to be at rest without losing generality 
  with normalised energy $\epsilon = m_{\rm p}/m_{\rm e}$. 
Correspondingly, the normalised energy of a target photon of frequency $\nu$ is given by  
  $\epsilon = h\nu/m_{\rm e}{\rm c}^2$.  
In photo-pion production calculations, 
  it is useful to define an invariant normalised energy, 
  $\epsilon_{\rm r} = \gamma_{\rm p}  (1-\beta_{\rm p}\mu) \epsilon$, 
  where $\mu$ is the cosine of the angle between the momentum vectors of the incident proton and the photon and $\beta_{\rm p}$ is the proton velocity normalised to the speed of light. 
These variables will be used hereafter.  

The length-scale over which a proton with energy $E_{\rm p}$ 
  would lose its energy through an interaction is 
\begin{equation} 
  r_{\rm int} = \left\{ \frac{{\rm d} \ln E_{\rm p}}{{\rm d} s} \right\}^{-1} 
    = \left\{  \frac{{\rm d} \ln \epsilon_{\rm p}}{{\rm d} s} \right\}^{-1} \ , 
\end{equation}      
   where ${\rm d} s$ is the differential path length element \citep[see][]{Blumenthal1970}.  
The energy loss length-scale may also be expressed in terms of the interaction timescale $t_{\rm int}$, i.e. 
 \begin{equation} 
   r_{\rm int} \approx {\rm c} \beta_{\rm p} t_{\rm int} \ . 
 \end{equation}
We adopt this expression in the calculations of the effective path length and the energy deposition rate 
  of CRs in our protogalactic model environments.  

In our calculations,  
  we restrict the particle energies to between $10^9$ and $10^{20}\;\!{\rm eV}$.  
These energies are sufficient to account for the relevant hadronic processes   
   in protogalactic environments. When modelling the CR production by SN events in protogalaxies, we further restrict the maximum particle energy to $10^{15}\;\!{\rm eV}$. This is a more realistic bound on the energies that could be attained by CRs accelerated in a SN remnant~\citep[e.g.][]{Schure2013MNRAS, Bell2013, Bell2013APh, Schure2013MNRAS, Kotera2011ARA&A, Bell1978MNRAS}, with higher energy particles likely to have an origin outside a typical protogalaxy~\citep[e.g.][]{Kotera2011ARA&A, Becker2008PhR, Blasi2014CRPhy, Hillas1984ARA&A}.  
   Below $10^9$ eV (i.e. 1 GeV), CRs are not thought to contain sufficient energy to drive a strong heating effect as they cannot undergo hadronic interactions below the GeV-level threshold energy.
Although the validity of this proposition is not strictly guaranteed,  
  taking an alternative approach does not imply more correct physics. 
 We lack concrete knowledge 
  of the properties of CRs outside our own solar system, 
  so those in the very furthest corners of the Universe are even less well known. 
We therefore omit unnecessary complications arising from the spatial variations and time evolution of cosmological CR spectra, 
  and adopt these working assumptions.

\subsection{Photo-pion Production}

\subsubsection{Interaction Channel \& Cross-Section} 

Photo-pion production arises when incident protons collide with the photons of a radiation field. 
The dominant interactions in photo-pion production are
 (i) resonant single-pion production, (ii) direct single-pion production, and (iii) multiple-pion production 
   \citep{Mucke1999PASA}. 
Other processes (e.g. diffractive scattering), though present, are less significant. 
The total cross-section of photo-pion production 
  is therefore the sum of the cross-sections of these three main interactions. 

Resonant single-pion production occurs through the production of $\Delta^+$ particles 
 which decay through two major channels,  
\begin{align}%
\label{eq:pg_int}%
\rm{p} \gamma \rightarrow \Delta^{+} \rightarrow%
	\begin{cases}%
	    \rm{p} \pi^0 \rightarrow \rm{p}2\gamma				\\[0.5ex]%
		\rm{n} \pi^+ \rightarrow \rm{n} \upmu^+ \nu_{\upmu}		\\%
		\hspace{4.1em} \myarrow \rm{e}^+ \nu_{\rm e} \bar{\nu}_{\upmu}%
	\end{cases}  %
\end{align}%
  \citep[see][]{Berezinsky1993}, 
  where charged and neutral pions are produced.  
The branching ratios for the $\Delta^{+}\rightarrow\pi^0$ and $\Delta^{+}\rightarrow\pi^+$ channels 
   are 2/3 and 1/3 respectively. 
Direct pion production is less efficient, 
  and its rate is roughly 1/3 of that of the resonant single-pion production at $\epsilon_{\rm r} \approx 500$. 
While single-pion production dominates at energies $\epsilon_{\rm r}$ below $\approx 3500$,  
  multi-pion production occurs at higher energies 
  \citep[see][]{Mucke1999PASA}.

Neutrons have a half-life of about $880\;\!{\rm s}$~\citep{Nakamura2010}. 
If not colliding with other particles, 
  the neutrons produced in the resonant single-pion process  
  will undergo $\beta^{-}$-decay, 
\begin{equation}%
\rm{n} \rightarrow \rm{p} \rm{e}^- \bar{\nu}_{\rm e} \ .  %
\end{equation}%
However, they may instead interact with the radiation field 
  which will lead to further $\rm p$ and $\pi^-$ production,  
\begin{align}%
\rm{n}\gamma \rightarrow \Delta^0 \rightarrow %
		\begin{cases} %
			\rm{p} \pi^-  \\[0.5ex]%
			\rm{n} \pi^0%
		\end{cases} \ , %
\end{align}%
with branching ratios BR$(\Delta^{0}\rightarrow\pi^-) = 1/3$ and BR$(\Delta^{0}\rightarrow\pi^0) = 2/3$ \citep{Hummer2010}.   
  
These branching ratios appear to imply that 
   there would be more neutral pions than charged pions in single-pion production processes.    
However, when taking the additional charged pions produced in the residual interactions into account
  (which include ${\rm p}\gamma \rightarrow \Delta^{++} \pi^{-}, \Delta^{0}\pi^{+}$), 
  each type of pion is found to be produced in roughly equal numbers 
   \citep[see][]{Dermer2009book}. 
  
Experimental data  \citep{Baldini1987} indicates that 
  the total cross-section of the photo-pion interaction 
  may be divided into three energy regions 
  of average cross-sections  
\begin{align}%
\sigma_{\rm \gamma \pi}(\epsilon_{\rm r}) \approx %
	\begin{cases}%
		340~\rm{\mu b} 	\quad\quad [\epsilon_{\rm th} \leq \epsilon_{\rm r} < 980]	\\[0.5ex]%
		240~\rm{\mu b} \quad\quad [980  \leq \epsilon_{\rm r} <  3500]	\\[0.5ex]%
		120~\rm{\mu b}		\quad\quad [ 3500 \leq \epsilon_{\rm r} ]%
            \end{cases} %
\end{align}%
\citep[c.f. the approximated two-step function used in][]{Dermer2009book}, 
  where $\epsilon_{\rm th} (\approx 390)$ is the threshold energy for the pion production process.  
The lower-energy incident protons will lose approximately 20\% of their energy 
  in the resonant single-pion production process, 
  with similar rates for the neutral and charged pion channels. 
  The higher-energy incident protons will lose approximately 60\% of their energy 
  in the multiple-pion production process \citep{Dermer2009book}. 
The effective inelastic cross-section is 
  $\hat{\sigma}_{\rm \gamma \pi} \equiv \sigma_{\rm \gamma \pi}(\epsilon_{\rm r})K_{\rm \gamma \pi}(\epsilon_{\rm r})$,   
with the inelasticity coefficient 
\begin{align}
K_{\rm \gamma \pi}(\epsilon_{\rm r}) \approx %
	\begin{cases}%
		0.2		\quad\quad [ \epsilon_{\rm th} \leq \epsilon_{\rm r} < 3500] \\[0.5ex]%
		0.6		\quad\quad [3500 \leq \epsilon_{\rm r} ]%
	\end{cases} \ .   %
\end{align}%
This yields
\begin{align} 
   \hat{\sigma}_{\rm \gamma \pi} \approx %
  \begin{cases}   
      68~\rm{\mu b} 	\quad\quad [\epsilon_{\rm th} \leq \epsilon_{\rm r} < 980]	\\[0.5ex]%
	48~\rm{\mu b} \quad\quad [980  \leq \epsilon_{\rm r} <  3500]	\\[0.5ex]%
	72~\rm{\mu b}		\quad\quad [ 3500 \leq \epsilon_{\rm r} ]%
 \end{cases} \ . %
\end{align}  
If we ignore the small variations in $\hat{\sigma}_{\rm \gamma \pi}$ over $\epsilon_{\rm r}$, 
  the effective cross-section may be expressed as 
$\hat{\sigma}_{\rm \gamma \pi} \approx  \hat{\sigma}_{\rm \gamma \pi}^*\;\! {\cal H}(\epsilon_{\rm r} - \epsilon_{\rm th})$, 
  where ${\cal H}(...)$ is the Heaviside step function. 
In our calculations we adopt $ \hat{\sigma}_{\rm \gamma \pi}^* = 70~\rm{\mu b}$ and 
 $\epsilon_{\rm th} = 390$, which are the same as the values 
    suggested by \citet{Dermer2009book} \citep[see also][]{Dermer2003}. 

\subsubsection{Energy Loss}

In a protogalactic environment, 
  the target photons for photo-pion production 
  are expected to be supplied by the stellar sources 
  and the cosmological microwave background (CMB).     
For the CMB, the radiation is locally isotropic and, in both cases, the spectrum is well modelled by a blackbody.  
  The photon number density is given by a Planck function: 
\begin{equation}%
\label{eq:planckphot}%
  n(\epsilon) = \frac{8\pi}{\lambda_{\rm C}^3}\;\! 
    \frac{\epsilon^2}{{\rm e}^{\epsilon\;\! m_{\rm e}{\rm c}^2/{\rm k}_{\rm B}T}-1} \ ,  %
\end{equation}%
   where $\lambda_{\rm C} = {\rm{h}}/m_{\rm e}\rm{c}$ is the electron Compton wavelength
    and $T$ is the effective temperature of the source.   
In an isotropic radiative field with a blackbody spectrum,
   the timescale for the photo-pion interaction is  
\begin{align}%
\label{eq:invtimescale_planck}%
t_{\rm \gamma \pi}(\gamma_{\rm p}) = %
	& \frac{2\gamma_{\rm p}^2 \lambda_{\rm C}^3}{8\pi \rm{c}}\left\{
	\int_{0}^{\infty} 
	   \frac{ {\rm  d}\epsilon }{{\rm e}^{\epsilon\;\! m_{\rm e}{\rm c}^2/{\rm k}_{\rm B}T}-1} 
	\int_{0}^{2\gamma_{\rm p} \epsilon} {\rm d} \epsilon_{\rm r}  \;\! 
	  \epsilon_{\rm r}\hat{\sigma}_{\rm \gamma \pi} \right\}^{-1}%
\end{align}%
  in the high-energy limit, i.e.\ $\beta_{\rm p} \rightarrow 1$~\footnote{Note that while this result is valid for the CMB, 
    stellar emission radiates from discrete points and requires a correction factor -- see section~\ref{sec:protogalaxies}.} 
     \citep{Dermer2009book}. 
Thus, the energy of the incident proton is deposited over a path length 
$r_{\rm \gamma \pi}  = {\rm c}\beta_{\rm p}  t_{\rm \gamma \pi}(\gamma_{\rm p})$. 
With $\beta_{\rm p} \approx 1$ and approximating the cross-section as 
  $\hat{\sigma}_{\rm \gamma \pi} \approx  \hat{\sigma}_{\rm \gamma \pi}^*\;\! {\cal H}(\epsilon_{\rm r} - \epsilon_{\rm th})$,
  we may evaluate the latter integral in equation \ref{eq:invtimescale_planck},  
  and by imposing the requirement that $2\gamma_{\rm p}\epsilon \geq \epsilon_{\rm th}$ 
  we obtain  
\begin{equation}%
\label{eq:photopion_losses}
r_{\rm \gamma \pi}   \approx  \frac{\lambda_{\rm C}^3}{8\pi \;\! \hat{\sigma}_{\rm \gamma \pi}^*}   
    \left( \frac{m_{\rm e}{\rm c}^2}{k_{\rm B}T}\right)^3 {\cal J}^{-1} \ , 
  \end{equation}%
where
\begin{equation} 
 {\cal J} =  \int_{\eta}^{\infty} {\rm d} \zeta \ \frac{\zeta^2 - \eta^2}{{\rm e}^{\zeta}-1} 
   = 2\sum_{k=1}^\infty \left(\frac{\eta}{k^2} +\frac{1}{k^3} \right) {\rm e}^{-k\eta} 
\end{equation}  
and 
 $\eta = \epsilon_{\rm th}m_{\rm e}{\rm c}^2  / 2\gamma_{\rm p} {\rm k}_{\rm B}T$.

\subsection{Photo-pair Production}

\subsubsection{Interaction Channel \& Cross-Section}

Leptons can be produced in photo-hadronic interactions through  
  the Bethe-Heitler process \citep{Bethe1934}
\begin{equation}%
\label{eq:BH1}%
{\rm A}\gamma \rightarrow {\rm A}' {l}^+ {l}^- \ ,%
\end{equation}%
   where ${\rm A}$ and ${\rm A}'$ are nucleons, 
   and ${l}^+$ and ${l}^-$ are positively and negatively charged leptons, respectively. 
Photo-electron pair production 
 \begin{equation}%
\label{eq:BH}%
\rm{p}\gamma \rightarrow \rm{p} \rm{e}^+ e^- %
\end{equation}%
 is a Bethe-Heitler process  
  and is the key contribution to photo-pair CR-energy losses \citep[see e.g.][]{Blumenthal1970, Klein2006} 
  over the range of energies of interest in this work.   
\citet{Stepney1983MNRAS} considered an analytic fit to the cross-section for photo-electron pair production  
   and obtained 
\begin{equation}%
\begin{split}
\hat{\sigma}_{\rm \gamma e}(\epsilon_{\rm r}) \approx  & %
	\bigg\{ \frac{7}{6\pi} \alpha_{\rm f}  
	\left( \psi - \frac{109}{42}  \right)  +\big[ 473.65 +241.26 \psi   \\ 
	&  +81.151 \psi^2 +5.3814 \psi^3 
	  \big] \left( \frac{10^{-5}}{\epsilon_{\rm r}} \right)     \bigg\}  \sigma_{\rm T} 
\end{split} 
\end{equation}%
 \citep[see also][]{Jost1950PR, Bethe1954, Blumenthal1970},   
 where $\alpha_{\rm f}$ is the fine structure constant, 
 $\sigma_{\rm T}$ is the Thomson cross-section, 
 and 
  $\psi = \ln 2\epsilon_{\rm r}$.  
For the parameter regime relevant to this study, $\epsilon_{\rm r} \gtrsim 60$, 
  and the expression for the cross-section above may be approximated as  
\begin{equation}%
\label{approx-sigma}
\hat{\sigma}_{\rm \gamma e}(\epsilon_{\rm r})\approx %
\left\{	\frac{7}{6\pi}\alpha_{\rm f} \ln\left[\frac{\epsilon_{\rm r}}{k_{\rm \gamma e}}\right] \right\} \sigma_{\rm T}
\end{equation}%
  with $k_{\rm \gamma e}$ taking a value of $\approx 6.7$. 
Without losing generality we fix $k_{\rm \gamma e} = 6.7$ in our calculations, 
   which we find to be appropriate for the energy range of interest.     
Note that the same expression for the cross-section (equation~\ref{approx-sigma}) was given in 
  \citet{Dermer2009book}, but $k_{\rm \gamma e} = 2$ (for $\epsilon_{\rm r} \gtrsim 40$) was considered instead.

\subsubsection{Energy Loss}

Assuming that the electron-positron pair produced in the photo-pair production process is formed 
  at rest in the zero-momentum frame of the interaction \citep{Dermer1991A&A} 
  and that the invariant energy of the interaction 
  is much larger than that of the target photon field such that the interaction energy is specified by the CR energy, 
  the timescale for interactions with a blackbody radiation field is 
\begin{equation}
\label{eq:photopair_inv_t}%
t_{\rm \gamma e} \approx 
 \frac{m_{\rm p}}{m_{\rm e}} 
  \frac{ \gamma_{\rm p}^2}{ \rm{c}} 
\left\{ \int_{\gamma_{\rm p}^{-1}}^{\infty} {\rm d}\epsilon\;\! 
   \frac{n_{\rm ph}(\epsilon)}{\epsilon^2}
    \int_{2}^{2\gamma_{\rm p}\epsilon} {\rm d} \epsilon_{\rm r}\;\! 
        \frac{\epsilon_{\rm r}\;\! \hat{\sigma}_{\rm \gamma e}(\epsilon_{\rm r})}{\sqrt{1+2\epsilon_{\rm r}}}  \right\}^{-1}
\end{equation}%
\citep[see][]{Protheroe1996APh, Dermer2009book},  
  which implies an interaction path length of
\begin{equation}%
\label{eq:photopair_losses}%
r_{\rm \gamma e}(\gamma_{\rm p}) \approx %
\frac{9}{112}
	\frac{m_{\rm p} \lambda_{\rm C}^3 \gamma_{\rm p}^3 b^{5/3}}{m_{\rm e} \alpha_{\rm f} \sigma_{\rm T}\mathcal{F}_{\rm \gamma e}} \hspace{0.2cm}~\rm{Gpc} \ ,%
\end{equation}%
where $b =  m_{\rm e} {\rm c}^2/\gamma_{\rm p} {\rm k}_{\rm B} T$ and 
\begin{equation}%
\mathcal{F}_{\rm \gamma e} = %
	\mathcal{C}(b) + \mathcal{D}(b)\ln\left[\frac{1}{0.974 b k_{\rm \gamma e}}\right]+ \left(0.382 b k_{\rm \gamma e}\right)^{3/2} \mathcal{E}(b) \ .%
\end{equation}%
The functions $\mathcal{C}(b)$,  $\mathcal{D}(b)$ and $\mathcal{E}(b)$ result from standard integrals, 
   with $\mathcal{C}(b) = 0.74$ and $\mathcal{D}(b) = \Gamma(5/2)\zeta(5/2)$ 
   when $b\ll1$, $\mathcal{C}(b) = b^{3/2}\ln(b)e^{-b}$ 
   and $\mathcal{D}(b) = b^{3/2}e^{-b}$ 
   when $b\gg 1$, and $\mathcal{E}(b) = -\ln[1-e^{-b}]$ for all values of $b$ \citep{Dermer2009book}. 
Note that equation~\ref{eq:photopair_losses} requires an additional factor in the case of a diluted radiation field 
   (see section~\ref{sec:protogalaxies}).

\subsection{${\rm pp}$ Pion Production}

\subsubsection{Interaction Channel \& Cross-Section}
\label{sec:pp_interaction_intro}

Like photo-pion production, ${\rm pp}$ pion production also produces charged and neutral pions 
   which subsequently undergo decays to deposit energy into their ambient medium. 
The interaction leads to the following dominant pion production channels
\begin{align}\label{eq:pp_interaction}%
\rm{p} + \rm{p} \rightarrow %
	\begin{cases}%
		&\rm{p}  \Delta^{+~\;} \rightarrow\begin{cases}%
				\rm{p} \rm{p} \pi^{0}  \xi_{0}(\pi^{0}) \xi_{\pm}(\pi^{+} \pi^{-}) \\[0.5ex]%
				\rm{p} \rm{p}  \pi^{+}  \pi^{-}  \xi_{0}(\pi^{0}) \xi_{\pm}(\pi^{+} \pi^{-}) \\[0.5ex]%
				\rm{p} \rm{n}  \pi^{+}  \xi_{0}(\pi^{0}) \xi_{\pm}(\pi^{+} \pi^{-})\\[0.5ex]%
			\end{cases} \\%
		&\rm{n} \Delta^{++} \rightarrow\begin{cases}%
				\rm{n} \rm{p} \pi^{+} \xi_{0}(\pi^{0}) \xi_{\pm}(\pi^{+} \pi^{-}) \\[0.5ex]%
				\rm{n} \rm{n} 2\pi^{+} \xi_{0}(\pi^{0}) \xi_{\pm}(\pi^{+} \pi^{-})\\[0.5ex]%
			\end{cases} \\%
	\end{cases} \ ,%
\end{align}%
where $\xi_{0}$ and $\xi_{\pm}$ are the multiplicities for neutral and charged pions respectively 
  and the $\Delta^{+}$ and $\Delta^{++}$ baryons are the resonances 
  \citep[see][]{Almeida1968, Skorodko2008EPJA}. The inelasticity, i.e. fraction of hadronic energy lost to pion production is around 0.6 (similar to the photo-pion interaction). The rest of the energy mainly goes into secondary hadronic particles which can interact further along their propagations. Appendix~\ref{sec:appendixb} demonstrates that almost 100\% of the original energy is transferred to pion production within just a small number of interactions.
  
The total inelastic cross-section is well described by the analytic parameterisation  
\begin{equation}%
\label{eq:pp_cs}%
\hat{\sigma}_{\rm p\pi} = \left( 30.7 - 0.96\ln(\chi) + 0.18(\ln\chi)^{2} \right)\left( 1 - \chi^{-1.9}   \right)^{3}~\rm{mb} 
\end{equation}%
\citep{Kafexhiu2014},
where $\chi =  E_{\rm p}/E^{\rm{th}}_{\rm p}$, and the threshold energy 
  $E^{\rm{th}}_{\rm p} = (2m_{\rm \pi^{0}}+m^{2}_{\rm \pi^{0}}/2m_{\rm p}) {\rm c}^2 \approx0.28~\rm{GeV}$ is the proton energy required for the production of a neutral pion through 
\begin{equation} 
   \rm{p}\rm{p} \rightarrow \rm{p} \rm{p}\pi^{0}  \ .
\end{equation} 
The branching ratios for the production of each pion species across all relevant channels can be estimated by considering their respective production cross-sections. We find that the parameterisations given by~\citet{BlattnigPRD2000} offer a reasonable fit to the data, accounting for single-pion and multi-pion production channels\footnote{We find there is a minor discrepancy between data at lower energies below 50 GeV and the~\citet{BlattnigPRD2000} parameterisations, but the difference is sufficiently small for our purposes.}. This gives branching ratios for the production of $\{\pi^{+}, \pi^{-}, \pi^{0}\}$ as $\{0.6, 0.1, 0.3\}$ at 1 GeV while, at higher energies, this levels out to $\{0.3, 0.4, 0.3\}$. This gives the total fraction of energy passed to neutral pions as around 0.3, while that going to charged pion production is around 0.7, of which 0.1 is lost to neutrinos~\citep{Dermer2009book}. The charged pions can further decay to electrons and further neutrinos by the weak interaction outlined in process~\ref{eq:weak_interaction2}. The neutrinos in this decay adopt around 75\% of the pion energy, while the remainder is passed to the electron and positron products and shared approximately equally between them according to their production multiplicity~\citep[see, e.g.][]{Lacki2013MNRAS, Lacki2012AIP, Loeb2006JCAP, Aharonian2000A&A}. Overall, this gives an inelasticity in the ${\rm pp}$ interaction of around 0.15 for the production of secondary electrons and positrons.\footnote{We acknowledge that our estimated inelasticity value is marginally lower than the 0.17 value quoted in~\citet{Aharonian2000A&A}, and implied in~\citet{Lacki2013MNRAS, Loeb2006JCAP}. We believe this is due to our additional inclusion of the 10\% losses to neutrinos in the charged pion production process.}

\subsubsection{Energy Loss}

The effective path length for CRs undergoing ${\rm pp}$ interactions is more straightforward than in previous cases. 
Before, the target field was comprised of photons, so a full relativistic treatment was necessary. 
Here, the target field is comprised of low-energy, non-relativistic baryons 
  that can be approximated as an ensemble of stationary particles. 
In this regime, the interaction path length is comparable to the classical definition of the mean free path
\begin{equation}%
   r_{\rm p\pi} = \frac{1}{\hat{\sigma}_{\rm p\pi} n_{\rm p}} \ ,%
\end{equation}%
   where $n_{\rm p}$ is the baryon number density in the target field, 
   and $r_{\rm p\pi}$ is the effective CR proton path length.

\section{The Protogalactic Environment}
\label{sec:protogalaxies}

Some protogalaxies exhibit strong, ongoing star-forming activity 
  \citep[see e.g.][]{DiFazio1979A&A, Solomon2005ARA&A, Pudritz2012, Ouchi2013ApJ, Knudsen2016MNRAS} 
  which inevitably results in a high rate of SN events.  
SN explosions generate shocks, providing a viable mechanism for particle acceleration 
   \citep[e.g.][]{Berezhko1999, Allard2007, Blasi2011, Bell2013, Schure2013MNRAS}. 
Frequent SN events inject substantial energy into their host galaxies 
  that can power a strong galactic wind outflow (cf. the nearby starburst galaxy M82).    
Shocks in such a galactic wind, like the SN-generated shocks,  
  can also serve as particle acceleration sites. 
  
Because of the larger spatial scales of galactic winds, CRs may achieve even higher energies than in SN shocks  
  if the magnetic field strengths in the two cases are similar \citep[see][]{Hillas1984ARA&A}.  
Moreover, star-forming activity and associated SN events would give rise to a population of young compact objects, 
  which are potential particle accelerators \citep[see e.g.][]{Dar2008PhR, Kotera2011ARA&A}. 
Protogalaxies are therefore conducive environments for particle acceleration, 
  and the CRs they produce would in turn affect the thermal properties of the ISM within the galaxy and local IGM. 

Here we adopt a generic model in which a protogalaxy is a parametric mean to specify  
  the baryon field, radiation field and magnetic field at a redshift of $z=7$. 
The complicated dynamical and internal star-forming processes  
  of the galaxies are ignored, and left to future work.  
Moreover, as an initial study, we assume that the CRs produced by local SN activity (assumed to be protons) 
  interact with only these three fields,  
  and the energy losses of the CRs is determined by their corresponding interaction processes.

\subsection{Density Field}
\label{sec:baryon_fields}

We model the (baryon) density field as an over-density on the background medium using the profile 
\begin{equation}
   \label{eq:density_profile}
   n_{\rm b}(r) = \frac{n_{\rm b, 0}}{\left(1+\xi_{\rm c}\right)\left(1+\xi_{\rm h}\right)}  
\end{equation}
  \citep{Dehnen1993MNRAS}, 
  where $\xi_{\rm c} = (r/r_{\rm c})^2$ and $\xi_{\rm h} = (r/r_{\rm h})^2$ for $r_{\rm c}$ 
  and $r_{\rm h}$ as the galaxy profile core and halo radius respectively, 
  as introduced in \citet{Dehnen1993MNRAS} and \citet{Tremaine1994}. 
This model is chosen for analytical tractability 
  while still being a reasonable description of the gas profile in an elliptical galaxy 
  for which the mass $M(<r)$ does not diverge with increasing radius, $r$. 
The core and halo radius are specified for each galaxy model. 
The values used for the parameters in this case are 
 $r_{\rm c} =1\;\! \rm{kpc}$ and $r_{\rm h} = 2\;\! \rm{kpc}$. 
The normalisation $n_0$ of the profile is $10\;\! {\rm cm}^{-3}$, 
  being a characteristic density of the protogalactic interstellar medium.

For the purposes of this study, the baryon profile is superposed onto a background density of $n_{\rm p} = 10^{-3}$ cm$^{-3}$. 
This fiducial value is taken as an example value to represent one possible intracluster medium environment 
  in which such a protogalaxy may be found. 
While a range of values either side of this would be reasonable for the various cluster environments and neighbourhoods 
  in which such a protogalaxy may be found, 
  we find the exact choice (if reasonable) bears little influence on the results of this study and focus of the discussion here.

\subsection{Radiation Field}
\label{sec:radiation_fields}

The radiation field plays two roles in this model. Firstly, as a target for the CRs in photo-pair and photo-pion interactions and, secondly, as a source of stellar radiation to drive a (conventional) heating effect with which the CR heating can be compared. In the first case, we must estimate the number density of target photons for CR interactions while, in the second case, the radiative stellar and X-ray flux is required.

\subsubsection{Radiation as a Cosmic Ray Target}

The spatial profile of the radiation field is governed by two factors: (1) the distribution of sources within the model protogalaxy and (2) the geometric spreading of photons as they radiate from their sources. In the case of a point source, the latter of these is described simply by the inverse square law, but for instances where the source is described more correctly by a distribution a more general result is required.

We assume that the stellar distribution in the galaxy follows the underlying \cite{Dehnen1993MNRAS} density profile 
    introduced in \S~\ref{sec:radiation_fields}, up to cutoff of $R = 10~{\rm kpc}$ above which we do not place any stars. 
   It should be noted that the vast majority of the stellar contribution is within $r_{\rm c} = 1~{\rm kpc}$ 
   \citep[see e.g.][]{Whalen2013ApJ, Mosleh2012, Ono2013} 
   and this high cut-off is intended to avoid a sharp unphysical edge to the galaxy.

Assuming that the radiative emission of the galaxy is dominated by that from stars, 
  and that each star may be modelled as a blackbody of temperature $T_{\rm *} = 30,000~{\rm K}$ 
  (i.e. type O/B), we can model the radiation field around a single source starting with the stellar luminosity, 
  given by the Stefan-Boltzmann relation as
\begin{equation}%
   L_{\rm *} = 4 \pi R_{\rm *}^2 \sigma T_{\rm *}^4 \ ,%
\end{equation}%
where $R_{\rm *}$ is the stellar radius. 
The energy density near a source is then
\begin{equation}%
   U_{\rm *} = \frac{L_{\rm *}}{4\pi r^2 {\rm c}} \ .%
\end{equation}%
The photon number density of the field is estimated by dividing the energy density by the modal photon energy. 
This is estimated as $E_{\rm ph} = 2.82\;\! {\rm{k}_{\rm B}} T_{\rm *} \approx 3\;\! {\rm{k}_{\rm B}} T_{\rm *}$, 
   being roughly the peak of the spectrum of the radiation field, hence
\begin{equation}%
   n_{\rm ph} \approx \frac{U_{\rm *}}{3 \;\! {\rm{k}_{\rm B}} T_{\rm *}} \ .%
\end{equation}%
The generalisation from a point source to a distribution is derived in Appendix~\ref{sec:appendixa},
   with the result given by equation~\ref{eq:result_radfield}. This gives the stellar photon number density profile as
\begin{align}%
\label{eq:spat_rad}%
   n_{\rm ph}(r)  \approx &\frac{L_* \cdot N \cdot \omega(r)}{4\pi r_{\rm gal}^2 {\rm c} \;\! k_{\rm B} T_{\rm *} \;\! D_{\rm N}} \ , 
\end{align}%
where $\omega(r)$ is further introduced as a weighting (in this case the normalised density profile), and with $D_{\rm N}=1+{N}^{-1/3}+{N}^{-2/3}$ for $N$ as the number of stars. We use $N=10^6$, a size of stellar population that provides a stellar luminosity consistent with a galaxy of supernova rate ${\cal R}_{\rm SN} \approx 0.1~\text{yr}^{-1}$, or star-formation rate ${\cal R}_{\rm SF} \approx 16~\text{M}_{\odot}~\text{yr}^{-1}$, in line with relations determined from, e.g. Starburst99 simulations~\citep[see, e.g.][for details]{Hirashita2003A&A, Leitherer1999ApJS}.

We may estimate the relative importance of losses due to proton-photon interactions with this radiation field by considering the effective interaction path lengths introduced in section~\ref{sec:interactions}. 
These do not account for the spreading of the radiation field due to the spatial distribution of the sources within the protogalaxy. 
Therefore, a scaling factor of ${n_{\rm A}}/{n_{\rm ph}}$ is applied to $r_{\rm \gamma \pi}$ and $r_{\rm \gamma e}$ 
   (equations~\ref{eq:photopion_losses} and~\ref{eq:photopair_losses} respectively), 
   where $n_{\rm A}$ is the photon number density at the surface of the radiation source 
   (essentially being the integrated blackbody spectrum, $n_{\rm A} = {8 \pi \Theta^3 \Gamma(3) \zeta(3)}/{\lambda_{\rm C}^3}$) 
   and $n_{\rm ph}$ is the estimated photon number density within the galaxy, using equation~\ref{eq:spat_rad} above. 
 $\Theta$ is defined by the temperature of the radiation source, in this case the stellar temperature, $T_{\rm *} = 30,000~{\rm K}$.

Additionally, the CMB permeates the protogalaxy. This is modelled as a spatially uniform blackbody of characteristic temperature determined by the redshift, as per equation~\ref{eq:planckphot}, 
  where $\Theta(z) = {{\rm k_{\rm B}} T(z)}/{m_{\rm e} \rm{c}^2}$ 
  and $T(z) = T_0 (1+z)$ is the redshift-dependent temperature of the radiation field 
  with $T_0=2.73~{\rm K}$ as the present CMB temperature \citep{Planck2016aA&A}. 
The resulting target radiation field relevant to the CR interactions is then the sum of both the stellar emission and the CMB.

\begin{figure}
	\includegraphics[width=\columnwidth]{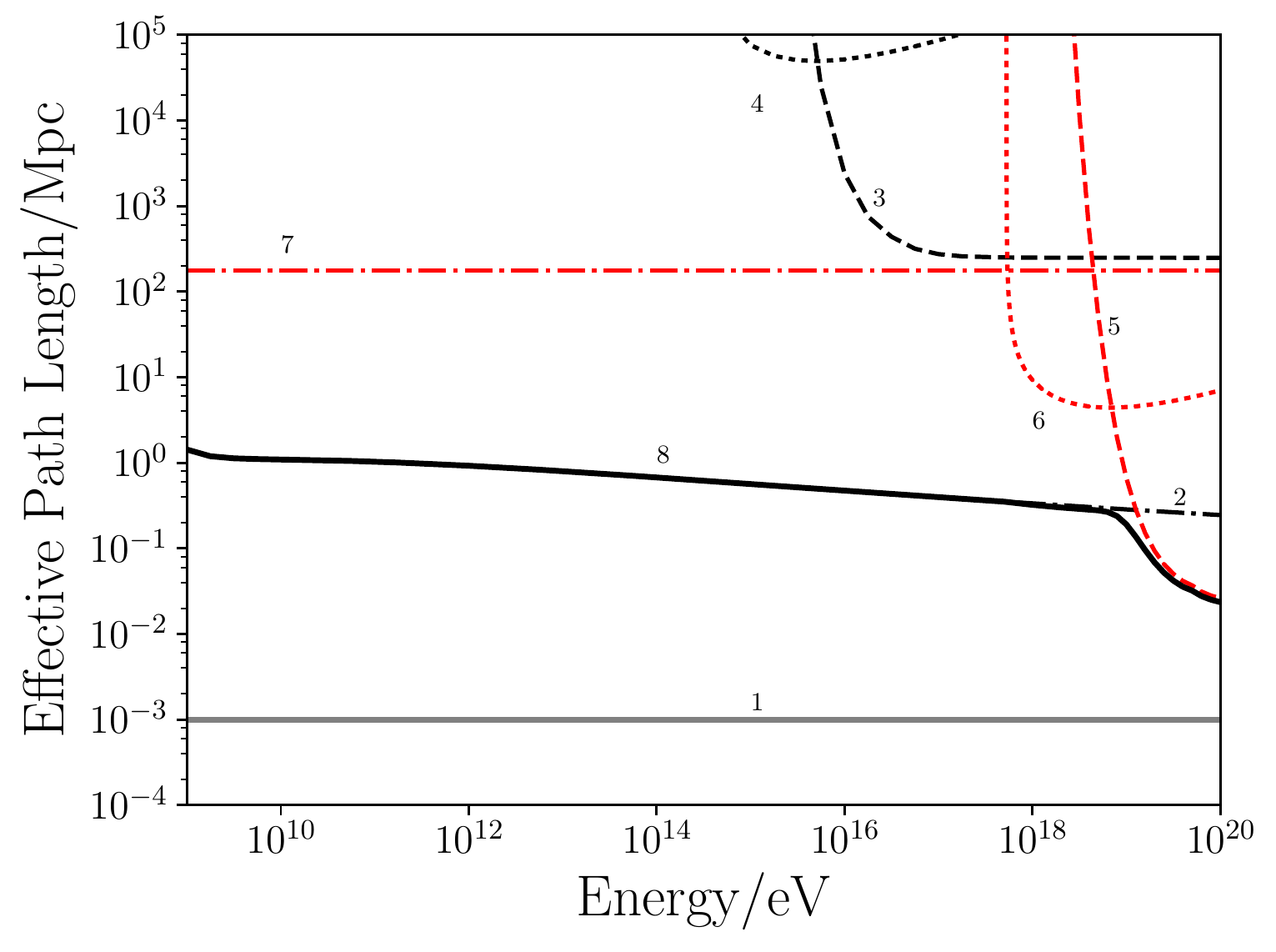}
    \caption{Plot to show the CR proton losses in terms of effective interaction path lengths. This results from a protogalaxy model with $n_{\rm p} = 10~\rm{cm}^{-3}$ and $N_{\rm *} = 10^6$, no galactic magnetic field and an underlying CMB radiation field at a redshift of $z=7$. The loss components due to the CMB are shown in red whereas those due to the foreground galaxy model are in black. The thick black line indicates the total effect, accounting for all energy loss contributions. The lines, as labelled, are (1) Protogalaxy Radius; (2) ${\rm pp}$ Pion-production (${\rm pp},{\rm p}\pi^{\pm}\pi^{0}$); (3) Galactic Radiation Photo-pion (${\rm p}\gamma,\pi^{\pm}\pi^{0}$); (4) Galactic Radiation Photo-pair (${\rm p}\gamma, e^{\pm}$); (5) CMB Photo-pion (${\rm p}\gamma,\pi^{\pm}\pi^{0}$); (6) CMB Photo-pair (${\rm p}\gamma, e^{\pm}$); (7) CMB Cosmological Expansion; (8) Total.}
    \label{fig:total_losses_b_0}
\end{figure}

We present in Fig.~\ref{fig:total_losses_b_0} the CR energy-loss components due to each of the interactions outlined in section~\ref{sec:interactions} in terms of their energy-dependent effective path lengths. For a protogalaxy such as this, it is apparent that the relative importance of the stellar photons is negligible, given that the associated path lengths (lines 3 and 4 respectively) are longer than those due to CMB interactions (red lines 5, 6 and 7). In turn, these losses are negligible compared to the interactions of CRs with the baryons of the galaxy (dashed black line 2) except at the very highest of energies, above 10$^{19}$ eV, at which point the CR flux is very low and thus such interactions are unlikely to bear any important influence on any resulting total energy deposition that arises. The thick black line indicates the resultant losses calculated as the reciprocal sum of all contributions, and this is almost completely governed by the ${\rm pp}$ interactions between UHECRs and the baryon (density) field of the galaxy, not the ${\rm p}\gamma$ interactions (photo-pair and photo-pion). Thus, the CR propagation lengths are strongly related to the protogalactic density field, with the radiation field bearing no significant influence on the resulting losses or any CR driven heating or energy-deposition effect.\footnote{In addition to the loss mechanisms introduced in earlier sections, the CRs free-streaming over cosmological distances 
  will undergo adiabatic losses due to the expanding Universe. 
The length-scale on which this arises is  
\begin{align}%
\label{eq:cosmological_losses}%
r_{\rm exp} &= {\rm{c}}\left|\frac{1}{E}\left(\frac{{\rm d}E}{{\rm d}t}\right)_{\rm exp}\right|^{-1}  = \frac{c}{{\rm{H_0}}\sqrt{\Omega_{\rm m}(1+z)^3 + \Omega_{\rm \Lambda}}} 
\end{align}%
\citep[see][]{Gould1975, Berezinsky1988, Berezinsky2006}, 
where $\rm{H_0} = (67.8\pm 0.9)$ km s$^{-1}$ Mpc$^{-1}$, $\Omega_{\rm m} = 0.308 \pm 0.012$ 
   and $\Omega_{\rm \Lambda} = 0.691 \pm 0.0062$ 
   with negligible curvature and radiation energy density contributions~\citep{Planck2016aA&A}.
These losses only become important for CRs that are not contained by the magnetic fields or interactions with their origin protogalaxy and are indicated in Fig.~\ref{fig:total_losses_b_0} by the dash-dotted horizontal red line (7).}
These loss lengths are calculated for the conditions at the centre of the model protogalaxy at $z=7$.

\subsubsection{Radiation as a Source of Heating}
\label{sec:rad_heating}

To provide a reference, we estimate the level of heating expected due to radiation from the stellar population and diffuse X-ray emission throughout the protogalaxy. To do this, we employ a Monte-Carlo (MC) method in which we distribute the total radiative emission across a simulated source distribution of 10,000 points. Accounting for the inverse square law and attenuation between each source point and some observation point $r$, the flux incident at a point $r$ due to each source may be found. The sum of this gives the incident flux at $r$ due to the entire source distribution. The level of heating is given by the product of the flux and the tendency of the medium at $r$ to absorb this radiation, quantified by the absorption coefficient $\alpha$ - itself a product of the local number density of scatterers and the scattering cross section of the radiation (likewise, the attenuation is determined by the integrated effect of this absorption along a photon's path). In the case of low-energy X-rays (we assume 1 keV), the cross section is approximately the Thompson scattering cross section, $\sigma_{\rm T}$, while for stellar emission we account for some energy dependence in that lower-energy non-ionising photons may additionally be scattered and attenuated by semi-ionised gas, trace metals and a trace level of dust (modelled to be present only in 30\% of the volume of interstellar medium, ISM\footnote{This is set to be the same as that estimated for the ISM of the Milky Way, corresponding to the volume fraction expected to be able to harbour trace dust and semi-ionised gas \citep[see, e.g.][]{Ferriere2001, Draine2011book}} where it is set to have 1\% of the average Milky Way dust fraction -- see section~\ref{sec:dust}) in the early ISM, 
 \citep[see][]{Cruddace1974ApJ} and ionising photons will scatter in a similar manner to the X-rays.

We assume that all non-ionising photons have energies equal to the peak energy in the blackbody spectrum ($E_{\rm \gamma} = 2.82~{k_{\rm B}} T_{\rm *}$) as this is the modal average energy of the photon gas. Considering separately the ionising radiation, we assume that these photons all have an energy of $13.6$ eV, the lowest energy at which they can be ionising -- this is within the Wein part of the spectrum, with most photons at the lower energies, close to the ionisation threshold. The fraction of the stellar radiation in the ionising and non-ionising part of the spectrum is estimated by numerically integrating the Planck function over energy in the two regimes.

While the luminosity of individual stars can be determined using the Stefan-Boltzmann law, as before, the X-ray emission from the galaxy is attributed to a range of sources and processes. 
  For simplicity, we assume a total galactic luminosity of $L_{\rm X} = 8.3\times 10^{38}\;\! {\rm erg~s}^{-1}$, corresponding roughly to a system with ${\cal R}_{\rm SN} \approx 0.1~\text{yr}^{-1}$ or ${\cal R}_{\rm SF} \approx 16~\text{M}_{\odot}~\text{yr}^{-1}$ according to scaling relation estimates in the literature~\citep[see, e.g.][]{Sarkar2016ApJ, Appleton2015ApJ, Li2013MNRAS}, and
    similar to the diffuse X-ray luminosity of the starburst M82~\cite[see, e.g.][]{Watson1984}. 
    We neglect point source contributions in our model because, at $z=7$, many source candidates would not have had sufficient time to form in abundance.
    This total X-ray luminosity is then distributed  evenly among the 10,000 MC simulation source points which,
     themselves are weighted by number according to the underlying density profile 
     from which radiative sources would be expected to form. 

The resulting heating effect of stellar photons within the central parts of the protogalactic ISM (where it is strongest) 
  is around $10^{-25}\;\!{\rm erg~cm}^{-3}{\rm s}^{-1}$ 
    while, for X-rays, the level is a little lower, around $10^{-26}\;\!{\rm erg~cm}^{-3}{\rm s}^{-1}$ -- in both cases, these rates correspond to ${\cal R}_{\rm SN} \approx 1.0~\text{yr}^{-1}$, or a star-formation rate of around ${\cal R}_{\rm SF} \approx 160~\text{M}_{\odot}~\text{yr}^{-1}$ (see conversion in section~\ref{sec:mag_field_devel}). 
 As the non-ionising stellar radiation is attenuated more effectively, 
    this heating rate falls below that due to X-rays outside the galaxy (see Fig.~\ref{fig:heating_sfr} for the full heating profile).

\subsubsection{Comparison with Previous Results}
\label{sec:radiative_comparisons} 

While previous works have examined radiative heating rates in protogalaxies, these are usually estimated in the context of balancing cooling rates when a system is assumed to be in thermal equilibrium~\citep[see, e.g.][]{Baek2005ApJ}. Specific studies regarding stellar and X-ray heating mechanisms in protogalaxies are not widespread, and the expected heating rates in such systems do not appear to have yet been established in the literature. In light of the lack of direct comparisons, we instead provide sanity checks of our radiative heating levels by comparison with 
 more local starburst and Galactic-type environments, which we scale appropriately.
 
 In evolved low-redshift galaxies, radiative heating (particularly UV) is reliant on the presence of dust and polycyclic aromatic hydrocarbons (PAHs) which absorb, scatter and reprocess the optical and soft UV radiation, eventually yielding an ISM heating effect~\citep[see e.g.][]{Serra2016ApJ, Meijerink2005A&A, Wolfire1995ApJ, Bakes1994ApJ, Hu2017MNRAS}. High-redshift protogalaxies, however, are considerably less abundant in dust and PAHs (see section~\ref{sec:dust}), which means these mechanisms do not efficiently operate. Instead, the main mechanism would be Thompson (Compton) scattering.
  
 In~\citet{Glover2012MNRAS}, heating rates for molecular clouds are considered when illuminated by a galactic-like interstellar radiation field (ISRF) as described in~\citet{Draine1978ApJS}, with cloud densities ranging from $10^1-10^6~\text{cm}^{-3}$. The relevant heating mechanisms are a combination of photo-electric heating by dust extinction, molecular hydrogen ($\text{H}_{2}$) photodissociation and the pumping of highly excited vibrational levels of $\text{H}_{2}$ by UV photons -- the `UV pump'~\citep{Burton1990ApJ}. In all instances, it is found that photo-electric heating dominates, with a power of around $10^{-25}~\text{erg cm}^{-3}~\text{s}^{-1}$. In our protogalaxy model, the ISM gas is assumed ionised, while the dust fraction is expected to be negligible compared to a local-Universe type galaxy, so none of the heating mechanisms explored in studies of low-redshift molecular clouds would be feasible dominant channels. Moreover, the ISRF would be substantially different, being more intense in a violently star-forming protogalaxy. 
 
 Nonetheless, the ~\citet{Glover2012MNRAS} result can be used as a basis for scaling between local and high redshift galaxy environments. This can be done by replacing the interaction cross section associated with the heating process, and scaling the ISRF to a level appropriate for high redshift galaxies. The cross-section of photo-electric grain heating is taken to be $1.5\times 10^{-21}~\text{cm}^2$~\citep[see][]{Draine1978ApJS}, about $10^4$ times larger than that for Thompson scattering.  The total ISRF luminosity in a protogalaxy is about $10^4$ times that of the Milky Way, derived from the star counts of the host galaxies ($10^6$ stars in the protogalaxy compared to $10^{11}$ stars in the Milky Way), the bolometric luminosities of those stars (type O/B stellar luminosity is around $10^6$ times greater than the low mass stars that tend to dominate the Milky Way population -- see~\citealt{Lada2006ApJ, Beech2011JRASC, Ledrew2001JRASC}) and the volume containing these stars ($10^3$ times smaller in the protogalaxy). Thus, in the case of pure Thompson scattering (reducing the cross section by a factor $10^4$), and in a protogalactic ISRF (increasing the intensity by a factor of around $10^4$), the stellar heating rate in the protogalaxy ISM would be roughly the same as that calculated in the molecular cloud, around $10^{-25}~\text{erg cm}^{-3}~\text{s}^{-1}$ - broadly in-line with what we find.

There are not many direct comparisons available in the literature for X-ray heating. However, the heating mechanisms are insensitive to the environments. The process of X-ray ionisation followed by thermalisation through Coulomb interactions of the emitted electron will arise in any partially-ionised medium. The scattering of the free-electrons (either in a plasma, or when electrons are emitted after ionisation processes) will also arise in systems where free-electrons are present. Complications such as dust scattering and metallicity are less important for X-ray heating than for stellar heating, and so rates in the literature are more broadly comparable to the results of this study. \citet{Meijerink2005A&A} finds a rate of $10^{-24}~\text{erg cm}^{-3}~\text{s}^{-1}$ for a characteristic starburst environment similar to the one considered in this paper, modelled on NGC 253, with $\mathcal{R}_{\rm SN} \approx 0.1~\text{yr}^{-1}$ \citep{Ferriere2001}, but for an ISM density of $10^3~\text{cm}^{-3}$. As the heating is determined by the density of the target ISM and intensity of the X-rays, it scales accordingly. Therefore, in a $10~\text{cm}^{-3}$ ISM, the X-ray heating power would be around $10^{-26}~\text{erg cm}^{-3}~\text{s}^{-1}$, which is roughly comparable to that found in section~\ref{sec:rad_heating}.
 
\subsubsection{Dust Scattering in Protogalaxies}
\label{sec:dust}

Starburst galaxies in the local Universe are rich in dust, presumably from sources such as Asymptotic Giant Branch (AGB) and evolved stars
  ~\citep[see e.g.][among others]{Ferrara2016MNRAS}. 
The build up of dust produced by AGB stars is expected to be on the evolutionary timescales 
  of the host galaxies 
  which, in some cases, is comparable to the Hubble time.  
Dust in protogalaxies in the early Universe is more likely to be 
  a product of SN events 
  resulting from the evolution of massive stars
  which occurs on a short timescale; 
  such candidates are core-collapse SNe \citep[see, e.g.][]{Gall2011A&AR, Bocchio2016A&A}. 
There are arguments that core-collapse SNe 
  might not be efficient in dust production, 
  as powerful shocks (particularly reverse shocks) in and around core-collapse SN remnants penetrate into the SN ejecta 
  and can destroy much of the dust that would otherwise form \citep{Bianchi2007MNRAS, Nozawa2007ApJ, Nath2008ApJ, Silvia2010ApJ}, 
  reducing the abundance considerably \citep[see, e.g.][]{Yamasawa2011ApJ}. 
Observations seem to support this, with evidence for considerably less dust in high-redshift galaxies, 
  particularly above $z\approx5$ 
   \citep{Bouwens2016ApJ, Capak2015Natur, Walter2012ApJ, Ouchi2013ApJ, GonzalezLopez2014ApJ, Ota2014ApJ, Riechers2014ApJ, Maiolino2015MNRAS}, 
  with the exception of quasar host galaxies~\citep{Beelen2006ApJ, Michalowski2010A&A}. 
A recent study \citep{Ginolfi2018MNRAS} pointed also to the evolution of the dust content in galaxies as they evolve -- 
   there could be a substantial reduction of dust content (by two orders of magnitude) 
  compared to a present-day galaxy (at $z = 0$) when tracing back to its very early stages, e.g. at $z\sim 7$ 
  and a significant reduction of dust content if the metal content in the environment is deficient. 
The strong scattering and radiative reprocessing effects by dust, 
  which could be important in local starbursts, e.g. M82~\citep[see, among others][]{Gao2015ApJ, Hutton2015MNRAS, Hutton2014MNRAS}, 
  would be unlikely to play an important role in reprocessing starlight 
  in metal poor protogalactic environments in the very early Universe 
  considered in this study. 
For this work, we therefore assume only a trace dust component in the host ISM. 

\subsection{Magnetic Field}
\label{sec:magnetic_fields}

Observations and simulations have indicated that young galaxies may have developed magnetic fields of strengths comparable to those in the current Universe within a few Myr of their formation 
  \citep{Bernet2008, Beck2012, Hammond2012arXiv, Rieder2016MNRAS, Sur2017}. 
This suggests that there must be mechanisms prevalent during the early protogalactic stages of galaxy evolution, in which magnetic field amplification mechanisms transformed seed magnetic fields 
  (of around $10^{-20}$ G, see e.g. \citealp{Sigl1997PRD, Howard1997}) 
  to the saturation magnetic fields of a few $\mu$G observed in galaxies today 
  \citep{Beck2005, Fletcher2011MNRAS, Adebahr2013, Beck2013}.

\subsubsection{Field Development in the Protogalaxy}
\label{sec:mag_field_devel}

A number of mechanisms have been proposed to drive this magnetic field growth, 
  most notably invoking turbulence driven by accretion and SN explosions of early generations of stars 
  \citep[see e.g.][]{Beck2012, Latif2013MNRAS}. 
\citet{Schober2013A&A} introduces an appropriate model 
   which demonstrates the evolution of the field according to the onset of turbulence 
   due to rapid star-formation and the resulting high SN rate \citep{Rees1987QJRAS, Balsara2004}. 
The model describes the time-evolution of the protogalactic magnetic fields on two scales: 
  a large-scale field, $B_{\rm L}(t)$, and a small-scale, viscous field $B_{\rm v}(t)$. 
On the viscous scale, the evolution of the magnetic field $\mathcal{Q}(t)$ is modelled 
   by an exponential increase of magnetic field strength from a seed field $B_0$ 
   (being the weak seed field permeating the protogalaxy 
   before the onset of star-formation of strength around 10$^{-20}$ G) 
   to the saturation strength field $B_{\rm v,sat} = \mathcal{S} (\ell_{\rm v}/L_{\rm f})^{\theta} f$ 
   at a growth rate $\Gamma$ 
   \citep[see][]{Braginskii1965RvPP, Kazantsev1968JETP} 
   in a time $t_{\rm v}$, i.e. $\mathcal{Q}(t) = B_0 \exp(\Gamma t)$, 
   after which the non-linear evolution begins to dominate. 
Note that $\theta$ is introduced here as the slope of the turbulent velocity spectrum, 
   which takes a value of around 1/3 in the case of incompressible turbulence~\citep{Kolmogorov1941}, 
   and increases with compressibility. Here, $\ell_{\rm v}$ is the viscous scale of the field, 
   $\mathcal{S} = \sqrt{4 \pi \rho}v_{\rm f}$ with $v_{\rm f} $ 
   as the typical fluctuation velocity of the largest forcing scale $L_{\rm f}$ sized eddies in the turbulent flow, 
   and $\rho = \rho(r)$ is the local protogalaxy matter density\footnote{Given our assumption
    that the magnetic field is predominantly driven by SNe, 
   it follows that its morphology should follow that of the SN population 
   which roughly traces the gas density profile of the galaxy, as accounted for in this term.}. 
In the case of SN-driven turbulence in a spherical protogalaxy model as considered here, 
   $v_{\rm f}  \approx R_{\rm gal} ({2}\pi \rho G/3)^{1/2}$ 
   and $L_{\rm f} \approx {\rho v_{\rm f} ^3}/{(\mathcal{R}_{\rm SF} \alpha_* \beta \xi \;\! \epsilon_{\rm T} E_{\rm SN})}$ \citep{Schober2013A&A}. 
   The velocity is estimated from a pressure/gravity equilibrium approximation 
   and the forcing scale $L_{\rm f}$ is estimated by comparing the energy dissipation rate with the energy input rate due to SNe 
   which is modelled to follow the star-formation rate $\mathcal{R}_{\rm SF}$ 
   via the relation $\alpha_* \beta\;\!  \xi \;\! \epsilon_{\rm T} \;\! \mathcal{R}_{\rm SF}/\rm{M}_{\rm SN}$. 
Here, $\epsilon_{\rm T} \approx 0.1$ is the fraction of energy transferred from the SN to turbulent energy~\citep[e.g.][]{Walch2015MNRAS},
and $\xi = 0.01$ accounts for the fraction of energy lost by the SN to neutrinos (see equation~\ref{eq:cr_scale} and following text for discussion of this parameter and its value).
 $\beta$ is introduced as an efficiency parameter specifying the fraction of SN energy transferred into ISM turbulence.  
 $\alpha_*$, which specifies the fraction of stars produced that ultimately yield a SN,      
  is determined by the initial mass function (IMF) of the stars which form in the galaxy 
  and also $M_{\rm SN}$, the mass of a star which eventually produces a SN event.  
It may be estimated as follows:  
\begin{equation}
    \alpha_* \approx \frac{\int_{M_{\rm SN}}^{M_{\rm max}} {\rm d}M\;\! M^{-\Upsilon}}
      {\int_{M_{\rm cut}}^{M_{\rm max}} {\rm d}M\;\! M^{-\Upsilon}}  \  ,   
\end{equation} 
where we use a value of $M_{\rm SN} = 8.5~\text{M}_{\odot}$ for the minimum mass progenitor mass for a core-collapse SN event (following e.g.~\citealt{Smith2011MNRAS})\footnote{This is consistent with a best-fit value of around $8\pm1~\text{M}_{\odot}$ -- see~\citet{Smartt2009MNRAS}, also~\citet{Smartt2009ARA&A}.}, and we adopt an upper limit for the progenitor mass of $M_{\rm max} = 50~\text{M}_{\odot}$\footnote{The upper limit for the progenitor mass is  uncertain and there is some debate in the literature. Values as low as $16.5~\text{M}_{\odot}$ are proposed in some cases~\citep[e.g.][]{Smartt2009MNRAS}. However, masses of core-collapse progenitor red super-giant stars are observed to exist up to masses of at least $25~\text{M}_{\odot}$~\citep[see][]{Smith2011MNRAS}. Moreover, SN Type IIn arise from stars with initial masses of above $40~\text{M}_{\odot}$~\citep[e.g.][]{Tominaga2008ApJ, Muno2006ApJ}, which pushes the core-collapse progenitor mass even higher~\citep[see also][]{Smith2004ApJ, Smith2009AJ}. Observational studies of M31 and M33 also suggest maximum cut-offs of around $35-45~\text{M}_{\odot}$~\citep[e.g.][]{Jennings2014ApJ, DiazRodriguez2018ApJ}. Since our model is not sensitive to the maximum value adopted for progenitor mass, we simply use the suggested theoretical maximum limit for a progenitor able to yield a SN event (above which direct collapse into a black hole is favoured), of $50~\text{M}_{\odot}$~\citep[see][]{Fryer1999ApJ, Heger2003ApJ}.}. To calculate the fraction of stars which yield SN events, we normalise over a mass range from $M_{\rm cut} = 1~\text{M}_{\odot}$ to $M_{\rm max} = 50~\text{M}_{\odot}$, which assumes a negligible population fraction of stars with $M>50~\text{M}_{\odot}$ (indeed, the power-law nature of the IMF suggests that the fraction of stars with $M>50~\text{M}_{\odot}$ would not be more than 1\% for any reasonable choice of maximum upper mass).
We otherwise use a Salpeter IMF with index $\Upsilon = 2.35$
    for deriving a conversion between the SN event rate ${\cal R}_{\rm SN}$ 
    and the star-formation rate ${\cal R}_{\rm SF}$. 
This gives $\alpha_* \approx 0.05$ and a conversion of
${\cal R}_{\rm SF} \approx 160~{\rm M}_\odot {\rm yr}^{-1} ( {\cal R}_{\rm SN}/{\rm yr}^{-1})$, which we use throughout this study.

The large-scale evolution of the magnetic field then follows the viscous evolution,
   scaled by $\left({\ell_{\rm v}}/{L_{\rm f}}\right)^{5/4}$ until a saturation time $t_{\rm v}$, 
   after which large scale turbulent growth dominates the evolution,
\begin{equation}
\label{eq:protob_L}
B_{\rm L}(r, t) = \begin{cases}
               \mathcal{Q}(r, t)\cdot\left(\frac{\ell_{\rm v}}{L_{\rm f}}\right)^{5/4}			
                   &\quad\quad [t < t_{\rm v}]				\\
               \mathcal{S}(r) \cdot f \cdot \left(\frac{\ell_{\rm p}(t)}{L_{\rm f}}\right)^{\theta + 5/4}  		
                    &\quad\quad [t_{\rm v} \leq t <t_{\rm sat}]	\\
               B_{\rm L,sat}(r)		&\quad\quad [t > t_{\rm sat}]		
            \end{cases} \ .
\end{equation}
Here, $\ell_{\rm p}$ is the scale of peak energy of the magnetic field 
  given by $\ell_{\rm v} + [v_{\rm f} (t-t_{\rm v})/L_{\rm f}^{\theta}]^{1/(1-\theta)}$
  \citep{Schober2013A&A},  
  and the saturation strength is $B_{\rm L,sat} (r) = \mathcal{S}(r) \;\! f$,
  where $f=0.1$ is  the fraction of the turbulent kinetic energy transferred to magnetic energy, 
  estimated from simulations 
  \citep[see][]{Federrath2011PRL}. 
  
\begin{figure}
	\includegraphics[width=\columnwidth]{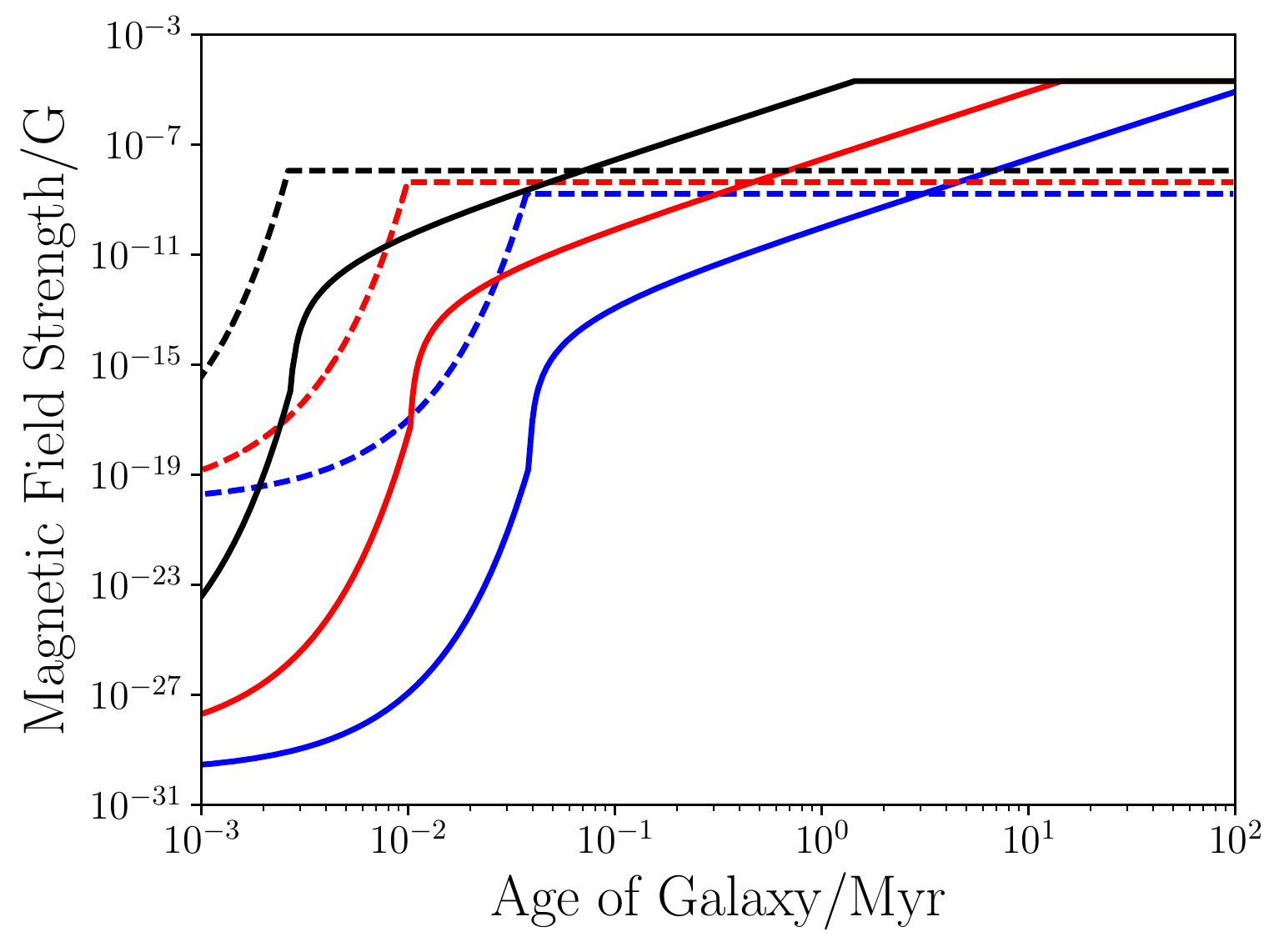}
    \caption{Magnetic field evolution 
         with SN rates of $\mathcal{R}_{\rm SN} = 0.1, 1$ and 10~${\rm yr}^{-1}$ (curves in blue, red and black respectively).  
     Solid lines indicate the evolution of the large-scale ordered magnetic field, 
     and dashed lines are the evolution for the viscous scale, turbulent magnetic field. 
   This uses values for $\alpha_* = 0.05$ and, based on simulations, $\beta = 0.05$ \citep{Latif2013MNRAS}.
   }
    \label{fig:bfield_evolution}
\end{figure}

We can use this formulation to model the magnetic field growth in a starburst protogalaxy 
  based on the star-formation rate (and hence the SN rate). 
While the star-formation rate in some protogalaxies is believed to be very high, 
  there is little agreement between different theoretical and observational studies 
  (particularly at the high redshifts of interest here) 
  with rates potentially varying from $0.1\;\!\ {\rm M}_{\odot}{\rm yr}^{-1}$ 
  in fairly quiescent cases to over $1000\;\! {\rm M}_{\odot}{\rm yr}^{-1}$ 
  in the most massive examples 
  \citep{Hernquist2003MNRAS, Robertson2010Nat, Ouchi2013ApJ, Stark2013ApJ, Barger2014ApJ}. 
 The rate of star-formation and the rate of CRs produced 
  are related through the rate of SN events.  
For local starburst galaxies, 
  the SN event rates are estimated 
  to be $\sim 0.1~{\rm yr}^{-1}$ for M82~\citep{Fenech2010MNRAS} and NGC~253~\citep{Lenc2006AJ} 
    and $\sim 4~{\rm yr}^{-1}$ for Arp 220~\citep{Lonsdale2006ApJ}. 
Assuming SN event rates ${\cal R}_{\rm SN}$ in the range between 0.1 to $10~{\rm yr}^{-1}$  
   would cover a reasonable fraction of plausible parameter space. Thus, we consider SN rates within this range 
  and assess the magnetic field development and the impact on CR-driven heating power.  
From the expressions for $v_{\rm f} $ and $L_{\rm f}$, 
  and by setting values for the efficiency parameters $\alpha_* = 0.05$ 
  and $\beta = 0.05$~\citep[based on simulations - see, e.g.][]{Latif2013MNRAS}, 
  we can specify models describing the development of the protogalactic magnetic field 
  for each SN rate, as per Fig.~\ref{fig:bfield_evolution}.

\subsubsection{Cosmic Ray Propagation}
\label{sec:CR_diff}
 
In a uniform magnetic field, the gyration of a charged CR particle is characterised by the Larmor radius, 
  which is given by   
\begin{equation}%
     r_{\rm L} = \frac{3.3 \times 10^{12}}{|Z|} \left(\frac{E}{10^{9}\;\!{\rm eV}}\right) \left( \frac{{\rm \mu G}}{B}\right) ~\rm{cm} \ , %
\end{equation}%
  where $Z$ is the charge of the particle.  
Propagation of CR particles in a medium permeated by a tangled magnetic field is more complicated. 
However, $r_{\rm L}$ can be used to derive a phenomenological description for CR transport in terms of diffusion, 
   in which the interaction the particles with the magnetic fields is treated as scattering.  
In this scenario the CR particles advect and diffuse spatially, 
  subject to radiative energy loss and momentum redistribution   
  until they collide with another particle 
  that results in a cascade of particle production. 
  
In this demonstrative study, we adopt this as our working scenario. 
Moreover, we consider that energy loss and momentum redistribution of the CR particles are insignificant 
  during particle propagation before the collision with another particle. 
In a galactic region without a large-scale flow, advective transport is also insignificant.  
Thus, the transport of CRs is described by a diffusion equation: 
\begin{equation}
\label{eq:diffusion_equation}
    \frac{\partial n}{\partial t} = \nabla \cdot \left[ D(E, {\mathbf r}, t) \nabla n \right] + Q(E, {\mathbf r}, t) \ ,
\end{equation}  
   where $n(E,{\mathbf r},t)$ is the number density of CR particles with energy $E$ at location ${\mathbf r}$ and time $t$.  
The two parameters in the equation are 
  the injection rate of the CR particles, 
  represented by the source function $Q(E, {\mathbf r}, t)$, 
  and the diffusive speed of the particles, 
  expressed in terms of the diffusion coefficient, $D(E, {\mathbf r}, t)$.     
For the case of spherical symmetry, only the radial component is relevant. 
Hence, the transport equation becomes 
\begin{equation}
\label{eq:diff2}
   \frac{\partial n}{\partial t} = \frac{1}{r^2}\frac{\partial}{\partial r}\left[D(E,r,t) ~ r^2 \frac{\partial n}{\partial r}\right] + Q(E, r,t) \ .  
\end{equation} 

Without losing generality we consider only CR protons. 
The diffusion coefficient takes the parametric form:   
\begin{equation}
   D(E, r, t) = D_0(t) \left[ \frac{r_{\rm L}(E, \langle B \rangle)\vert_{r,t}}{r_{{\rm L},0}(t_{\rm sat})}\right]^{1/2}
\end{equation}  
   where $\langle B \rangle$ is a characteristic mean magnetic field strength, evaluated at a location $r$ and time $t$, 
  and $t_{\rm sat}$ is the time at which the galactic magnetic field has evolved to reach a saturation state.    
Here, the reference Larmor radius $r_{{\rm L},0}(t_{\rm sat})$ 
   corresponds to that of a 1 GeV CR proton gyrating around a 5$\mu$G uniform magnetic field,  
   and the normalisation $D_0(t_{\rm sat}) = 3.0\times 10^{28}$ cm$^2$ s$^{-1}$, 
   a value roughly equal that which is observed in the Milky Way ISM 
   \cite[see, e.g.][]{Berezinskii1990, Aharonian2012, Gaggero2012}. 
   
The source function is a product of two separable components: 
\begin{equation}
   Q(E, r; t) = \bigg\{\mathcal{L}_{\rm CR}(E)\;\! S_{\rm N}(r)\bigg\}\bigg\vert_t \ ,   
\end{equation} 
  with the energy spectrum of the CR particles given by $\mathcal{L}_{\rm CR}(E)$, 
  and the rate of the CR injection per unit volume represented by $S_{\rm N}(r)$.
As stochastic acceleration generally gives a power-law energy spectrum, 
    $\mathcal{L}_{\rm CR}(E)$ assumes the form 
\begin{equation}
    \mathcal{L}_{\rm CR}(E) =  \mathcal{L}_{0} \left(\frac{E}{E_0}\right)^{-\Gamma} \ . 
\end{equation}     
We set the power-law index $\Gamma = 2.1$,  
   following observations of the galactic differential CR flux power spectrum
      (see \citealt[][]{Kotera2011ARA&A, Kotera2010JCAP, Allard2007}, although we acknowledge that steeper indices of 2.3--2.4 have been suggested more recently in the case of pure proton spectra, e.g.~\citealt{HESSA&A2018, Adrian-Martinez2016PhLB}).  
Note that this index corresponds to a flatter spectrum than that of CR electrons 
  which suffer more severe radiative losses \citep[see e.g.][]{Amenomori2008}. 
Using the lowest energy under consideration in this work, $10^9$~eV, as the reference energy $E_0$ 
   and the maximum energy of interest for CR injection, $E_{\rm max} = 10^{15}~{\rm eV}$, 
   we define the normalisation 
\begin{equation}
\label{eq:cr_norm}
   \mathcal{L}_0 = \frac{L_{\rm CR, eff}(1-\Gamma)E_0^{-\Gamma}}{E_{\rm max}^{1-\Gamma} - E_0^{1-\Gamma}}  \ , 
\end{equation}
   where $L_{\rm CR, eff}$ is the power in the CR protons. 
Supposing that the CR protons are consequential of SN events, 
 then $L_{\rm CR, eff}$ may be expressed as   
\begin{equation} 
\label{eq:cr_scale}
   L_{\rm CR, eff} = \varepsilon \xi E_{\rm SN} {\cal R}_{\rm SN} 
     =   \alpha_* \left[\frac{\varepsilon \xi E_{\rm SN} \mathcal{R}_{\rm SF}}{M_{\rm SN}}  \right]   \ .  
\end{equation}
  where ${\cal R}_{\rm SN}$ is the SN event rate,  
     $E_{\rm SN}$ is the total energy generated per SN event, 
     $M_{\rm SN}$ is the mass of the SN progenitor star, 
     $\mathcal{R}_{\rm SF}$ is the star-formation rate,  
     $\alpha_*$ is the fraction of stars to yield a SN,  
     $\varepsilon$ is the fraction of SN power converted into to CR power 
     and $\xi$ accounts for the fraction of SN energy lost to neutrinos.   
We assume that $E_{\rm SN} \approx 10^{53}$~erg 
  (the energy output of core-collapse Type II P SNe and hypernovae,  
  that have a massive low-metallicity progenitor),  
   $M_{\rm SN} \sim M_{\rm SN, max} \approx 50~{\rm M}_\odot$ (i.e. the maximum realistic mass of a SN event to attain a conservative estimate on the CR luminosity), $\varepsilon \sim 0.1$\footnote{
   Recent observational and theoretical studies suggest a range between 7\%~\citep{Lemoine2012A&A} and 30\%~\citep{Caprioli2012JCAP, Fields2001A&A} would be appropriate for this parameter, with 10\% (i.e. $\varepsilon \sim 0.1$) usually being taken as a characteristic value~\citep[e.g.][among others]{Dermer2013A&A, Morlino2012A&A, Strong2010ApJ, Wang2018MNRAS}. In line with this, we choose $\varepsilon \sim 0.1$, a conservative estimate within a realistic range of parameter values.} and $\xi \sim 0.01$\footnote{There are great uncertainties of in these parameters. 
  In particular,  $\xi$ depends the types of SNe expected 
    and the ISM environment  
\citep[see e.g.][]{Iwamoto2006}. 
  Core-collapse SNe have massive progenitors, 
   and most of their energy is carried away by neutrinos.    
 $\xi$ is expected to be around 0.01, although some have argued that it could be as low as 0.001 
    \citep[see reviews, e.g.][]{Smartt2009ARA&A, Janka2012ARNPS}. 
 For Type 1a SNe, which are less likely to arise in protogalaxies populated by massive young stars, 
    $\xi$ would be reflective of neutrino losses of a few percent~\citep[e.g. as per models and simulations in][]{Wright2017PhRvD}.}.

The following computational scheme is adopted to determine the CR particle density in our model galaxies 
   and how they diffuse outward. 
CRs are injected as discrete packets,  
  and the total CR power is the sum of all injection episodes.  
If there are $N_{\rm S}$ injection packets in total,   
  the spectral energy density of each is simply 
\begin{equation}
\label{eq:injection_function}
   N_{\rm CR}(E) = \frac{\mathcal{L}_{\rm CR}(E)}{N_{\rm S}} \ . 
\end{equation}      
With this prescription, 
   each injection episode is an independent event 
   and hence it serves as the initial condition for the homogeneous diffusion equation,   
\begin{equation}
\label{eq:diff3}
   \frac{\partial n}{\partial t} = \frac{1}{r'^2} \frac{\partial} {\partial r'}\left[ D(E, r', t) ~r'^2 \frac{\partial n}{\partial r'}\right]      
\end{equation} 
  (with $r'$ as the radial distance from the injection location ${\mathbf r}_{\rm s}$), 
  governing the evolution of the CR particles after their injection.  
The diffusion equation with a position-dependent coefficient $D(E, r', t)$ 
  can be solved using the method of Greens' functions.  

For the case of a single, discrete injection of duration $\Delta t$ initiated at $t = 0$, 
  equation~\ref{eq:diff3} gives 
\begin{equation}
\label{eq:diff_sol1a}
   n(E,r',t) = \frac{N_{\rm CR}(E)
     \vert_{{\mathbf r}_{\rm s}}\;\! \Delta t}{\left[4 \pi D(E,r',t)\;\! t \right]^{m/2}}\exp{\left\{-\frac{r'^2}{4 D(E,r',t)\;\! t}\right\} } \ . 
\end{equation} 
The geometrical parameter here $m = 3$, 
  which corresponds to an injection at a point source and the subsequent diffusion through an infinite volume\footnote{For a point source in an infinite plane $m=2$, and for a point source on a line $m=1$. 
  The protogalaxy that we model in this paper is not a point source 
     but a collection of sources spherically distributed within. 
These sources do not interact with each other, 
     and $m=3$ is applied in deriving the contribution from each individual.}.  
The solution for injections at multiple episodes  
  can be obtained by a convolution of the solutions  
  for the individual instantaneous injections.   
In the continuous limit, the solution\footnote{Obtained by Wolfram Mathematica.} is   
\begin{equation}
    n_{\rm T}(E,r',t) = \frac{N_{\rm CR}(E) \vert_{{\mathbf  r}_{\rm s}}\;\!t ~\Gamma(a, b)}
      {(1+p)\left[4\pi D(E,r',t)\;\! t \right]^{m/2}} \left\{\frac{r'^2}{4 D(E,r',t)\;\! t}\right\}^{\Upsilon}  \  ,  
\label{eq:diff_sol4}
\end{equation}
where the index is given by 
\begin{equation}
   \Upsilon = \frac{1}{1+p}-\frac{m}{2}  \ , 
\end{equation}
    and the arguments in the upper incomplete Gamma function $\Gamma(a, b)$ are
\begin{equation}
   a = \frac{-2 + m + p m}{2+2p}
\end{equation}
and
\begin{equation}
   b = \frac{r'^2}{4  D(E,r',t)\;\! t} \  . 
\end{equation}
The parameter $p$ is the index for the late-time evolution of the non-saturated magnetic field, $\propto\;\! t^p$. 
  This depends on $\theta$, the slope of the turbulent velocity spectrum -- see \S~\ref{sec:mag_field_devel}. 
A value of $p=2.88$ is appropriate here, 
  but $p=0$ after the magnetic field growth has saturated.

\section{Cosmic Ray Heating in Protogalaxies}
\label{sec:discussion}

\subsection{Heating Rate}

  
  \subsubsection{Microphysics of Energy Deposition}
  \label{sec:microphysics}
  
  Cosmic ray heating effects are driven by specific interaction channels. For example, $\gamma$-rays and neutrinos are relatively weakly interacting, and are unlikely to contribute significantly to local energy deposition or ISM heating. Electrons (also including positrons) are much more instrumental in driving a heating effect and are produced in the decays of secondary charged pionic particles.
  
  The secondary pions produced in the ${\rm pp}$ process inherit a large fraction of the primary CR energy over a small number of interactions. In a single interaction, secondary pions directly inherit around 60\% of the CR primary energy, with the rest leading to secondary neutrons and protons which can undergo further interactions. It can be shown that within just a small number of events of these secondary neutrons and protons, nearly 100\% of the primary energy is transferred to pions (see Appendix~\ref{sec:appendixb}) so that instantaneous energy transfer to to secondaries at the site of the initial interaction is a reasonable approximation. 
    
  The charged pion secondary particles decay to form electrons and positrons (we refer to both of these together as simply `electrons' for simplicity here) which inherit around 25\% of the pion energy~\citep[with the rest going into neutrinos, see, e.g.][]{Lacki2013MNRAS, Lacki2012AIP, Loeb2006JCAP, Aharonian2000A&A, Dermer2009book}. A single electron-yielding interaction will generally produce more than one electron and, at the mean energy of the CR spectrum used in our model (according to the power-law distribution described in section~\ref{sec:CR_diff}) ${\bar E} \approx 2.5~{\rm GeV}$ gives a secondary electron multiplicity of around 4 (when rounded to the nearest integer)\footnote{This value is determined empirically by fitting functions to experimental $\text{pp}$-interaction multiplicity data from~\citet{Breakstone1984PhRvD, Slattery1972PRL, Thome1977NuclPhys, Ansorge1988ZPhys, Rimondi1993proc, Alner1985PhLB, Alexopoulos1998PhLB, Albajar1990NuPhB, Arnison1983PhLB, Alner1984PhLB, Abe1990PhRvD, Whitmore1974PhR, Wang1991thesis}. The review paper~\citet{Fiete2010JPhG} and analysis in~\citet{Albini1976NCimA} find a function of the form $a+b s^{\mathcal{W}}$ provides a good description of the multiplicity data, where $s$ is the centre-of-mass interaction energy (in GeV) and the parameters take values of $a = 0.0$, $b=3.102$ and $\mathcal{W}=0.178$~\citep{Fiete2010JPhG}.}. As such, the energy transferred from the CR primary to the electrons must be split equally among these secondary particles. For our calculations, we assume the average multiplicity of 4 is suitable for all energies, as the lower energies of the power law distribution of comparable multiplicities dominate the interactions. The higher multiplicities associated with higher energy CR interactions do not compensate for the greatly lower abundance of the primaries initiating the secondary cascades when adopting a spectral index of -2.1.
  
  The energy passed to each of the secondary electrons is therefore around 3.75\% of the CR primary energy (i.e. 0.6 of the energy is passed to charged pions, of which 0.25 is passed to the electrons, and this is split between an average of around 4 particles per CR primary), meaning a 1 GeV CR primary would ultimately yield 4 secondary electrons, each with energies of around 40 MeV. These electrons can then shed their inherited energy via three principal processes: bremsstrahlung, Coulomb interactions (and/or ionisation in cool, dense partially ionised plasmas) and radiative cooling, e.g. inverse-Compton scattering off CMB and starlight photons, or synchrotron cooling in the interstellar magnetic fields. Of the radiative cooling mechanisms, inverse-Compton scattering off the CMB and/or starlight dominates in a protogalaxy ISM due to the higher energy density of these radiation fields ($U_{\rm CMB} \approx 1,100 ~\text{eV cm}^{-3}$ at redshift $z=7$, compared to $U_{*} \approx 150 ~\text{eV cm}^{-3}$, according to equation~\ref{eq:radresult} with $N = 10^6$ and $N\;\!L_{*} \approx 2.8\times 10^{44}~\text{erg s}^{-1}$ for a galaxy with $\mathcal{R}_{\rm SN} = 0.1~\text{yr}^{-1}$, and scaling in proportion to SN event rate -- see section~\ref{sec:energetics}) compared to that of the magnetic fields (even when saturated at around $20~\mu\text{G}$, $U_{\rm B} \approx 12 ~\text{eV cm}^{-3}$).

The cooling rate due to inverse-Compton scattering with the CMB or starlight takes the form 
\begin{equation}
\dot{E}_{\rm rad} = \frac{4}{3}\sigma_{\rm T} {\rm c} \left(\frac{E_{\rm e}}{m_{\rm e} {\rm c}^2}\right)^2 U_{i}
\label{eq:rad_cool_elec}
\end{equation}
per particle \citep[see, e.g.][]{Blumenthal1970, Rybicki1986book} with $E_{\rm e}$ as the energy of the secondary CR electron, and where $U_{i}$ is the energy density of the radiation field, either the CMB, $U_{\rm CMB}$, or starlight, $U_{*}$ (synchrotron cooling takes the same form per particle, with the magnetic field energy density simply substituted for the radiation field energy density). The rate due to bremsstrahlung (free-free emission) per particle may be written as
\begin{equation}
\dot{E}_{\rm ff} \approx \alpha_{\rm f} {\rm c} \sigma_{\rm T} n_{\rm p} E_{\rm e} \,
\label{eq:ff_cool_elec}
\end{equation}
\citep[see, e.g.][]{Schleicher2013A&A, Dermer2009book} where $\alpha_{\rm f}$ is introduced as the fine structure constant, 
while that due to Coulomb interactions with the ISM (if modelling it as a low-density fully-ionised plasma) is given by
\begin{equation}
\dot{E}_{\rm C} \approx m_{\rm e} {\rm c}^2 n_{\rm p} {\rm c} \sigma_{\rm T} \ln \Lambda \ ,
\label{eq:c_cool_elec}
\end{equation}
\citep[see, e.g.][]{Schleicher2013A&A, Dermer2009book} where $\ln \Lambda\simeq 30$ is the Coulomb logarithm accounting for the ratio between the maximum and minimum impact parameters.
The relative importance of these processes can be assessed by considering their associated timescales,
\begin{equation}
\tau  = \left|\frac{1}{E_{\rm e}}\left(\frac{{\rm d}E_{\rm e}}{{\rm d}t}\right)\right|^{-1} \ ,
\end{equation}
which can be used to estimate the fraction of energy lost by the CR secondary electron through the respective process. 

\begin{figure}
	\includegraphics[width=\columnwidth]{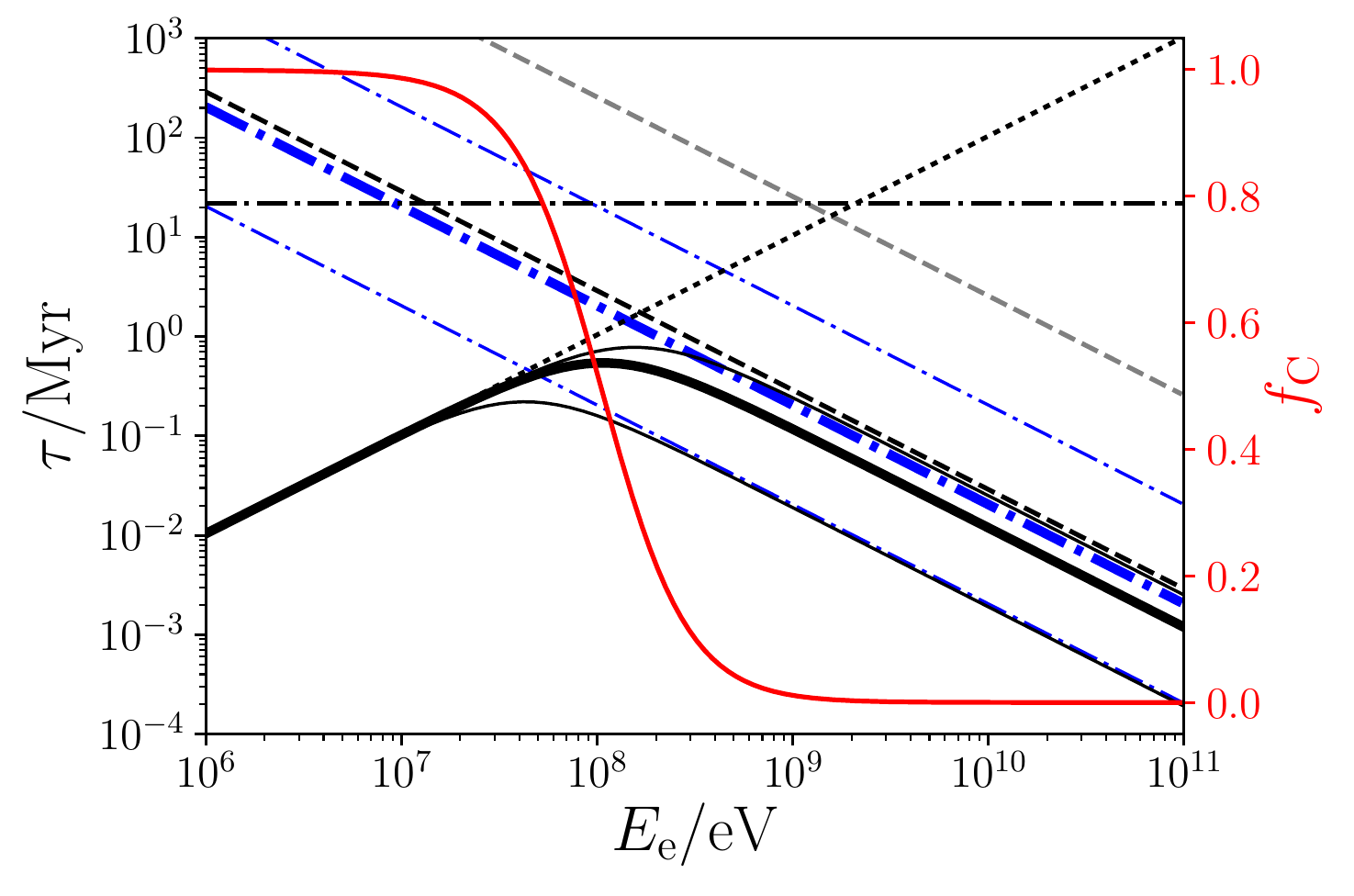}
\caption{Energy loss timescales for CR secondary electrons with energy $E_{\rm e}$: Coulomb (black dotted line), radiative inverse-Compton losses with the CMB (black dashed line) and stellar radiation fields (blue dashed lines -- the thickest due to the radiation field resulting from a protogalaxy model with $\mathcal{R}_{\rm SN} = 1~\text{yr}^{-1}$, the upper line for $\mathcal{R}_{\rm SN} = 0.1~\text{yr}^{-1}$ and the lower line for $\mathcal{R}_{\rm SN} = 10~\text{yr}^{-1}$), radiative synchrotron in saturated protogalactic magnetic field (dashed grey line), free-free bremsstrahlung (black dotted-dashed line). The solid black line shows the total energy loss timescales, the thick line in the case of the stellar radiation field with $\mathcal{R}_{\rm SN} = 1~\text{yr}^{-1}$ and thin lines for the other stellar radiation field models. The solid red line shows the thermalisation efficiency, $f_{\rm C}$, which is the fraction of CR electron energy transferred by the Coulomb-driven thermalisation process to the ISM compared to other processes. Below $E_{\rm e} = 10^7~\text{eV}$ this is approximately 1.0, indicating almost full thermalisation while, above $E_{\rm e} = 10^9~\text{eV}$, $f_{\rm C}$ becomes very small, indicative of inefficient thermalisation with most CR electron secondary energy lost to other processes. This assumes an ISM density of $n_{\rm p} = 10~\text{cm}^{-3}$.}
\label{fig:timescale_v_energy}
\end{figure}

\begin{figure}
	\includegraphics[width=\columnwidth]{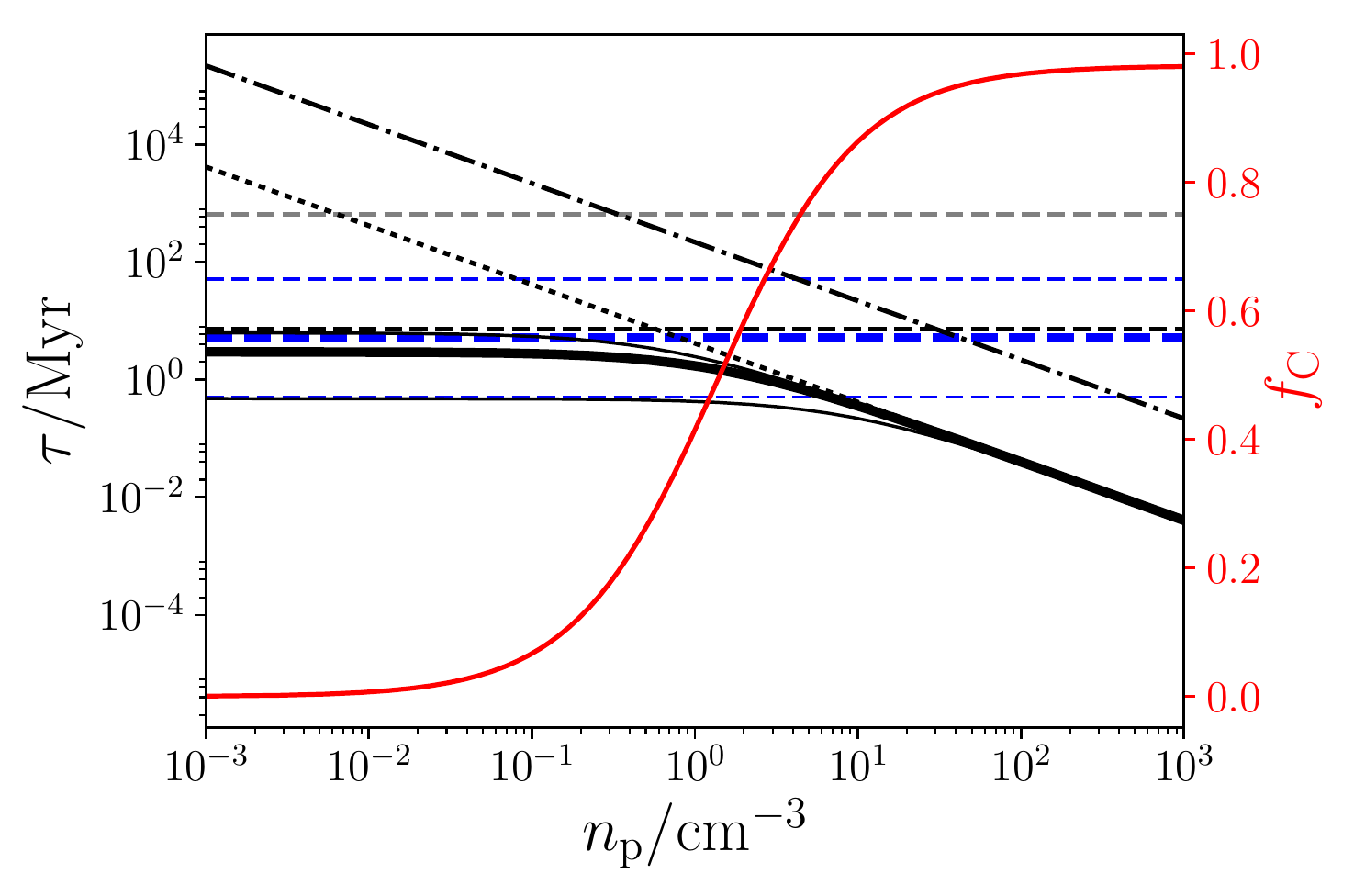}
\caption{Energy loss timescales for CR secondary electrons over different ISM densities:
Coulomb (black dotted line), radiative inverse-Compton losses with the CMB (black dashed line) and stellar radiation fields (blue dashed lines -- the thickest due to the radiation field resulting from a protogalaxy model with $\mathcal{R}_{\rm SN} = 1~\text{yr}^{-1}$, the upper line for $\mathcal{R}_{\rm SN} = 0.1~\text{yr}^{-1}$ and the lower line for $\mathcal{R}_{\rm SN} = 10~\text{yr}^{-1}$), radiative synchrotron in saturated protogalactic magnetic field (dashed grey line), free-free bremsstrahlung (black dotted-dashed line). The solid black line shows the total energy loss timescales, the thick line in the case of the stellar radiation field with $\mathcal{R}_{\rm SN} = 1~\text{yr}^{-1}$ and thin lines for the other stellar radiation field models. The solid red line shows the thermalisation efficiency, $f_{\rm C}$, which is the fraction of CR electron energy transferred by the Coulomb-driven thermalisation process to the ISM compared to other processes. Below $n_{\rm p} = 10^{-2}~\text{cm}^{-3}$ the value of $f_{\rm C}$ is very small, suggesting CR heating is very inefficient. Above $10^{1}~\text{cm}^{-3}$, it is approximately 1.0, indicating almost full thermalisation. This assumes a CR secondary electron initial energy of $E_{\rm e} = 40~\text{MeV}$, approximately corresponding to a CR primary proton energy of around 1 GeV.}
\label{fig:timescale_v_density}
\end{figure}

In terms of thermalisation, the ISM heating by the electrons is predominantly driven by Coulomb interactions. 
The energy of the CR primary ultimately passed into ISM heating is given by the product of $f_{\rm e}$ (being the fractional branching ratio associated with the formation of charged electron secondaries) and $f_{\rm C}(E_{\rm e})$, being the fraction of energy lost by secondary electrons to heating the ISM through Coulomb processes, given by
\begin{equation}
f_{\rm C}(E_{\rm e}) = \frac{\tau_{\rm C}^{-1}}{\tau_{\rm C}^{-1} + \tau_{\rm rad}^{-1} + \tau_{\rm ff}^{-1}}\Bigg\vert_{E_{\rm e}}\ ,
\label{eq:heat_fraction}
\end{equation}
with $\tau_{\rm C}$, $\tau_{\rm rad}$ and $\tau_{\rm ff}$ as the Coulomb, radiative (inverse-Compton and synchrotron) and free-free (bremsstrahlung) cooling timescales for the CR secondary electrons respectively. The total fraction of CR primary energy $E$ which is thermalised into the ISM then follows as $f_{\rm therm}(E) = (\xi_{\rm e}~f_{\rm C}~f_{\rm e}) |_{E_{\rm e}}$ for $\xi_{\rm e}$ as the electron production multiplicity (we use $\xi_{\rm e} = 4$ and $E_{\rm e} = 0.0375 E$). The energy dependence of the timescales (plotted in black) and the resulting impact this has on the Coulomb thermalisation fraction (i.e. $f_{\rm C}$ - plotted in red) is shown in Fig.~\ref{fig:timescale_v_energy}, where the timescales and thermalisation fractions are shown in terms of the CR secondary energy, $E_{\rm e}$. Here, it is apparent that the lower energy secondaries (arising from 1-10 GeV primaries) are much more important in driving ISM heating process than higher energy particles for which the losses are largely dominated by the radiative processes which are less directly involved in heating the local environment. Equations~\ref{eq:rad_cool_elec} to ~\ref{eq:c_cool_elec} also indicate a dependency on the local matter density (i.e. on the environment), and this is further explored in Fig.~\ref{fig:timescale_v_density} where, again, timescales are shown with black lines, and the Coulomb thermalisation fraction is shown in red. Here, it follows that higher densities provide a much more effective heating target, while CRs are less efficiently thermalised by this mechanism in lower density environments. 

\subsubsection{Individual sources}

As CR particles are highly relativistic, their speeds approach the speed of light, ${\rm c}$.   
The rate of heating per unit volume 
  by the particles through hadronic interactions at a location $r'$ from an individual source is 
\begin{equation}
    H(r')\vert_t =  \int_{E_{\rm min}}^{E_{\rm max}} {\rm d}E ~\mathcal{A}(r')~n_{\rm T}(E,r',t)\;\! {\rm c}~ \alpha(E, r')  \ ,   
\label{eq:heating_calc} 
\end{equation} 
i.e. with $n_{\rm T}(E,r',t)\;\! {\rm c}$ giving the rate at which CRs pass through a surface at $r'$.
The absorption coefficient $\alpha (E,r')$ 
    is the sum of the contributions from all interaction channels of all target particle species, i.e.  
\begin{equation}
   \alpha(E, r') = \sum_{\rm x} f_{\rm C}(E) n_{\rm x}(r') \sigma_{\rm x}(E)
   \label{eq:absorption_coefficient}
\end{equation}
  where $n_{\rm x} (r')$ is the number density of target particles of species ${\rm x}$ 
  and $\sigma_{\rm x}(E)$ is their corresponding interaction cross-section. $f_{\rm C}(E)$ results from the microphysics considerations in section~\ref{sec:microphysics}, and $\mathcal{A}(r')$ is a mean attenuation term resulting from the absorption of CRs along their propagations (see section~\ref{sec:mc_simulation} for details).
  
\subsubsection{Monte-Carlo Simulation}
\label{sec:mc_simulation}

Equations \ref{eq:diff_sol4} and \ref{eq:heating_calc} determine the heating contributed by a single CR source. 
The sources are independent, 
  and hence the heating at a location is the linear sum of the contributions of all sources in the system. 
This formalism lends itself efficiently to computing the heating using a Monte-Carlo method. 

We consider a weighted Monte-Carlo scheme in which a total number of $N_{\rm S}$ sources are generated, 
  with their spatial distribution weighted according to the density profile of the protogalaxy 
  up to its characteristic edge (set to be 1~kpc).
The size of an individual source is set to be 0.01~kpc,  
  cf. the characteristic size of SN remnants \citep[see, e.g.][]{Badenes2010}. 
This, together with the efficiency of CR production scaled according to the star-formation rate, 
  provides the initial condition and hence the normalisation for equation~\ref{eq:diff_sol4}.   
We adopt a value of $N_{\rm S} = 10^4$ which, 
  we find, yields an acceptable signal-to-noise ratio in the resulting simulated CR density profile. 
The spatial density distribution of the target baryons involved in calculating the hadronic interactions 
  follows the density profile of the galaxy. 
  
The absorption of CRs along their propagations due to ${\rm pp}$ interactions is encoded in equation~\ref{eq:heating_calc} by the term $\mathcal{A}(r')$. While the attenuation experienced by free-streaming radiation can be calculated according to the optical dept, $\tau_{\rm free}$, that of CRs undergoing diffusive propagation through their attenuating medium takes a modified form. While the CRs propagate on a microscopic level at the speed of light, ${\rm c}$, their macroscopic propagation appears to be much slower than this due to the strong magnetic scattering they experience. As such, they will encounter a much greater amount of attenuating material over a given distance than freely streaming particles and the effective CR optical depth must be adjusted accordingly, i.e. $\tau_{\rm CR}^*(r') = \psi(r') \tau_{\rm free}(r')$ for $\psi(r')$ as the scale factor. These quantities are calculated locally: i.e. indicating the optical depth experienced by a particle propagating through an infinite medium with the properties at the point $r'$.

In the free-streaming limit, the timescale associated with absorption due to the ${\rm pp}$ interaction is
\begin{equation}
t_{\rm p\pi} = \frac{\ell_{\rm p\pi}}{\rm c} = \frac{1}{{\rm c}~n_{\rm p}~\hat{\sigma}_{\rm p\pi}} \ .
\label{eq:pp_path_scale}
\end{equation}
Over such a timescale, a particle in the diffusion limit would macroscopically be displaced by a characteristic distance
\begin{equation}
\ell_{\rm diff} = \sqrt{4~D~t_{\rm p\pi}} = \sqrt{\frac{4~D}{c~n_{\rm p}~\hat{\sigma}_{\rm p\pi}}} \ ,
\end{equation}
although, when considering the microphysical scatterings, its actual path length would be the same as $\ell_{\rm p\pi}$ in equation~\ref{eq:pp_path_scale} -- just compressed into a much smaller region due to the scattering to give a reduced macroscopic displacement. It therefore follows that the same level of attenuation would be experienced (i.e. with the same optical depth) after a particle has propagated a length $\ell_{\rm p\pi}$ in the free-streaming regime, or $\ell_{\rm diff}$ in the diffusion regime. The ratio between these path lengths therefore gives the level of additional attenuation that would arise in the diffusion regime compared to a free-streaming particle travelling in a straight line, i.e.
\begin{equation}
\psi(r') = \frac{\ell_{\rm p\pi}}{\ell_{\rm diff}} = \sqrt{\frac{{\rm c}}{4~D~n_{\rm p}~\hat{\sigma}_{\rm p\pi}}} \ .
\end{equation}
As such, the attenuation can be calculated in the free-streaming limit, and the optical depth scaled by $\psi(r')$ to estimate the attenuation experienced by the CRs in the diffusion regime. The calculation itself is done by considering the path lengths between the heating position $r'$ and each of the sources in the ensemble of $N_{\rm S}$ MC sources. The free-streaming attenuation along each line of sight is then estimated and the adjusted by $\psi$ such that the resulting attenuation factor is
\begin{equation}
\mathcal{A}(r') = \exp \left[-\tau_{\rm CR}^*(r')\right] = \exp \left[-\psi(r') \tau_{\rm p\pi}(r')\right] \ .
\end{equation}

\subsubsection{Results}
\label{sec:results}

\begin{figure}
  \vspace*{0.1cm}
	\includegraphics[width=\columnwidth]{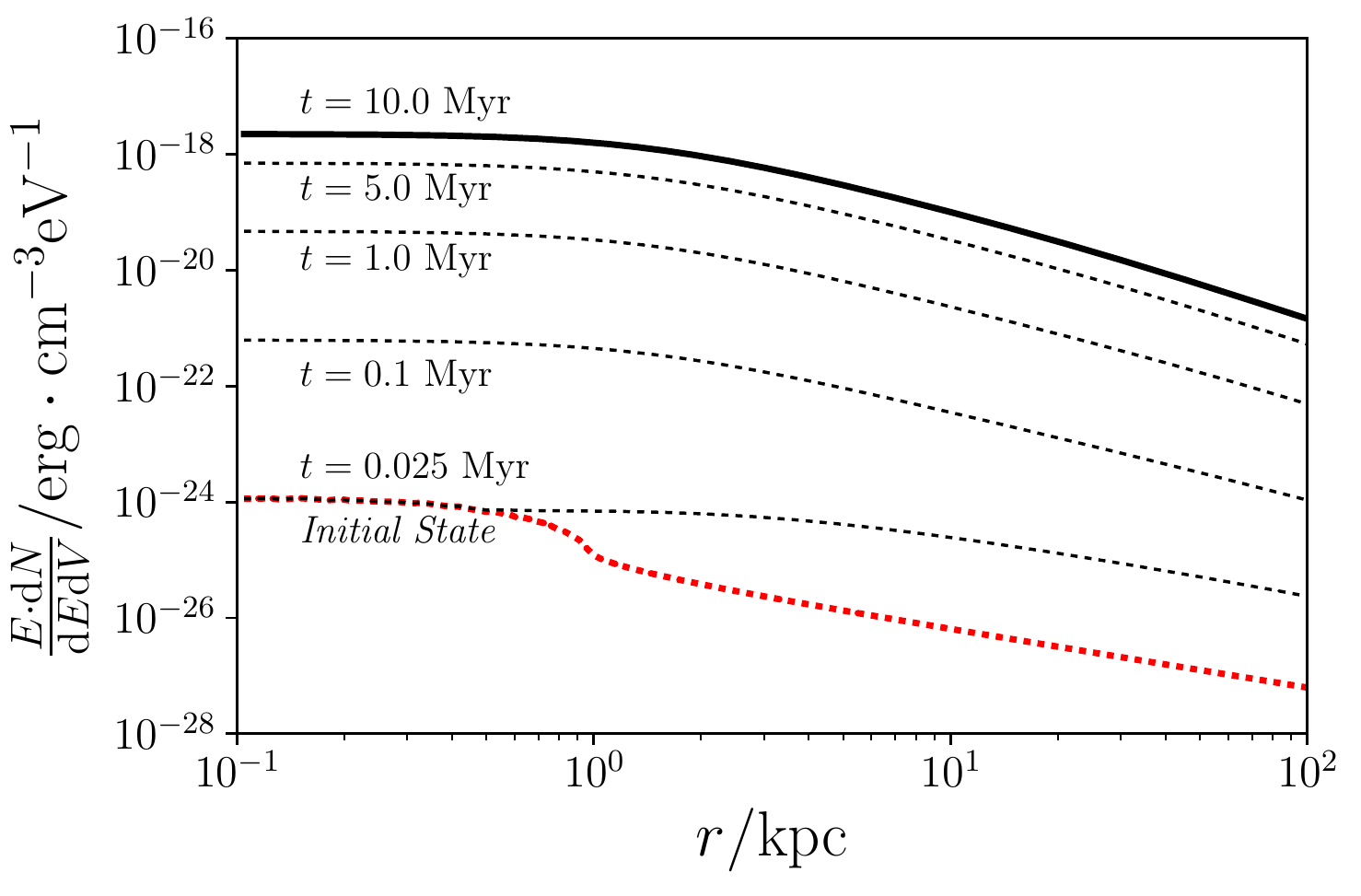}
    \caption{The evolution of the CR profile (at $\bar{E} \approx 2.5$~GeV) with the protogalactic magnetic field 
       for SN rate ${\cal R}_{\rm SN}  = 1.0\;\! {\rm yr}^{-1}$ (i.e. our intermediate model in terms of SN rate). 
      The red dashed line corresponds to the initial free-streaming state, 
             while the solid black line corresponds to the steady state profile
                in which the magnetic field evolution has saturated (by around 10~Myr).    
      The black dashed lines correspond to the interim stages at $t = 0.025, 0.1, 1.0$ and $5.0$~Myr, as labelled.}
    \label{fig:diffusion_evolution}
\end{figure}

We consider protogalaxies that would be present at redshift $z \approx 7$. 
These represent the most distant galaxies to be observed in the near future, 
   e.g. in the ultra-deep field of the Subaru HSC deep-field survey \citep{subaru_proposal}.    
Fig.~\ref{fig:diffusion_evolution} shows 
  the evolution of the CR density distribution (at $\bar{E} \approx 2.5$~GeV) in the protogalaxy when 
  adopting a uniform SN event rate ${\cal R}_{\rm SN} = 1~{\rm yr}^{-1}$. 
Initially, the galactic magnetic field is not strong enough to confine the CRs,   
  and the energetic charged particles freely stream across the galaxy  
  until they hit a target particle in the interstellar medium (ISM), 
  or the intergalactic medium (IGM), at which point they are destroyed.  
After a Myr or so, containment of CRs sets-in as   
   the magnetic field in the galaxy gains substantial strength. 
After around 10.0~Myr, 
  the magnetic field evolution has reached saturation  
  and the CR density profile converges to a steady-state.  
As the CR particle number density 
  is linearly scaled with the SN event rate 
  (equation~\ref{eq:cr_scale}),   
  the normalisation of the steady-state profile (in Fig.~\ref{fig:diffusion_evolution})
  is also linearly scaled with the SN event rate 
  (and hence also the star-formation rate) of the host galaxy.

\begin{figure} 
   \vspace*{0.1cm}
	\includegraphics[width=\columnwidth]{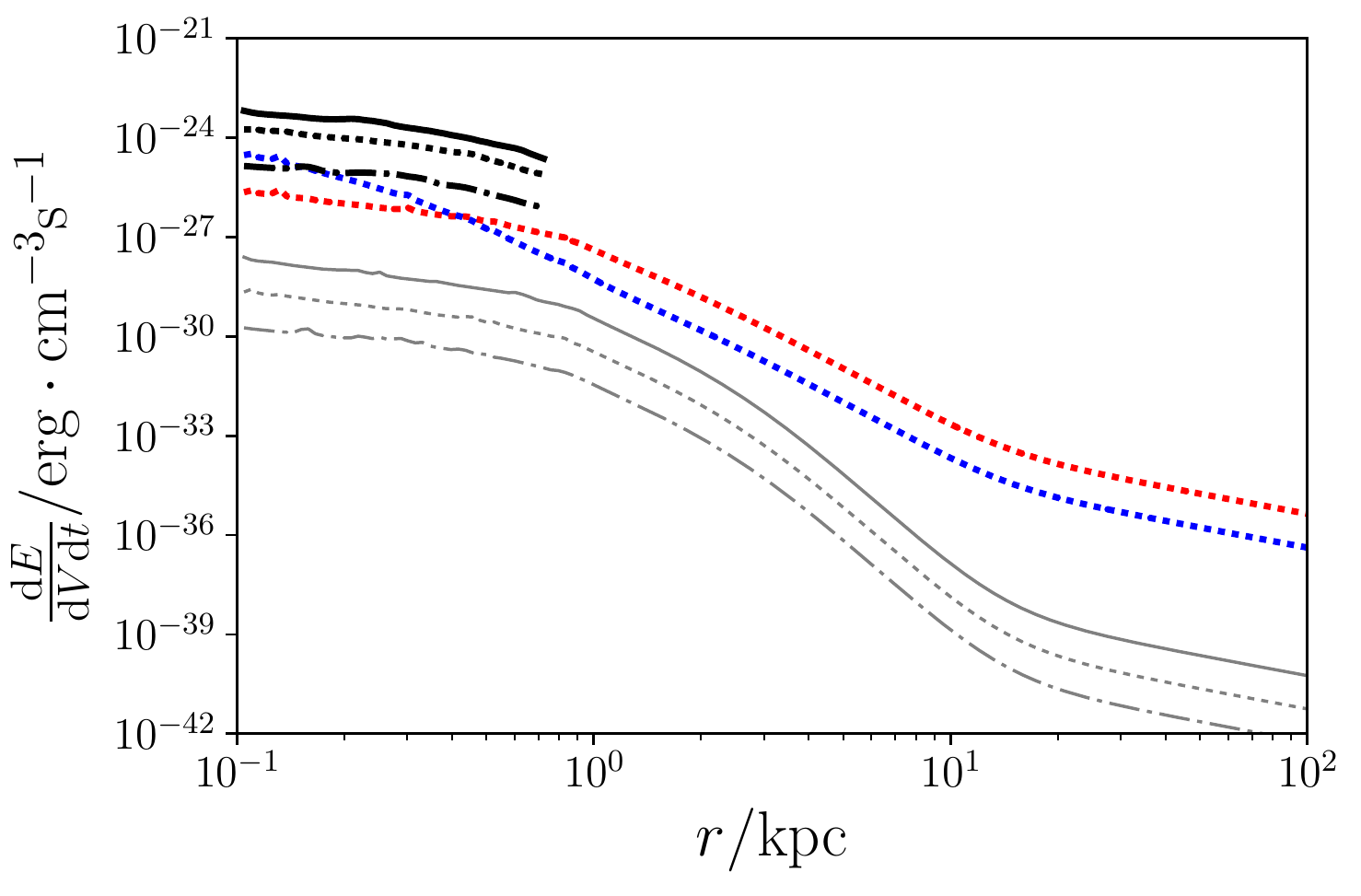}
    \caption{CR heating profiles for SN event rates at $\mathcal{R}_{\rm SN} = 0.1, 1$ and $10~{\rm yr}^{-1}$ 
      (dot-dashed, dotted and solid lines respectively) before (in grey) and after (in black) containment of the CRs 
        by the galactic magnetic field. 
       The profiles for radiative heating by starlight and X-rays are shown in blue and red respectively, corresponding to ${\cal R}_{\rm SN} = 1.0~\text{yr}^{-1}$, or to a star-formation rate of around $160~{\rm M}_\odot {\rm yr}^{-1}$ by the relation ${\cal R}_{\rm SF} \approx 160~{\rm M}_\odot {\rm yr}^{-1} ( {\cal R}_{\rm SN}/{\rm yr}^{-1})$ (see section~\ref{sec:mag_field_devel}). 
    For CR heating in the presence of magnetic field confinement, 
        only the profile segments for the region within the inner parts of the galaxy are shown.  
    The CR heating rate is expected to fall back down to values similar to those of the free-streaming cases 
        in regions outside the galaxy at distances beyond 10~kpc. }
\label{fig:heating_sfr}
\end{figure} 
 
We calculate the heating power of CRs in these galaxies and assess them as CR calorimeters 
as their galactic magnetic fields evolve \citep[see][]{Lacki2011ApJ, Thompson2007}.   
Fig.~\ref{fig:heating_sfr} shows the profiles of the CR heating power 
  for three SN event rates (${\cal R}_{\rm SN} = $ 0.1, 1 and 10 ${\rm yr}^{-1}$) 
  in the initial (free-streaming) stage 
  and the later steady-state stage when the evolution of the galactic magnetic field has saturated.   
For comparison, 
   we also compute the heating power due to starlight and diffuse X-rays 
   as per the prescriptions described in \S~\ref{sec:rad_heating}, where lines correspond to models with ${R}_{\rm SN} = 1.0~\text{yr}^{-1}$, or to a star-formation rate of around $160~{\rm M}_\odot {\rm yr}^{-1}$ by the relation ${\cal R}_{\rm SF} \approx 160~{\rm M}_\odot {\rm yr}^{-1} ( {\cal R}_{\rm SN}/{\rm yr}^{-1})$ (see section~\ref{sec:mag_field_devel}).
For the initial stage, 
  the CR heating profiles are shown spanning from the inner region to beyond the the galaxy. 
For the steady-state stage, only the profile segments 
   for the region within the ISM of the protogalaxy are shown. 
In this region CR transport is facilitated mainly by diffusion.   
Far beyond this central region, the CR particles are practically free-streaming,  
  and the heating rate level is expected to drop off to approximately that of the initial stage.
  The transitional region in the outskirts of galaxy is less straightforward to model.  
The CR transport and heating effect depends strongly 
  on the morphological structure of the magnetic field.    
The time-dependent magnetic activity in the region and the interface between the magnetic 
ISM and the magnetic IGM is not well understood, so the magnetic field properties in these galactic outskirts are very uncertain.  
Furthermore, the formulation of a kinetic description of particles 
  in the transition from a diffusion regime to a free-streaming regime (as would arise in these outskirt regions) 
  is theoretically challenging.  
It deserves an independent study in its own right,  
  and so we leave this to a more thorough future investigation. 

The development of the magnetic field leads to the suppression of the free-streaming propagation of CRs, 
   and the subsequent particle confinement leads to an amplification of the CR heating effect.   
Without CR containment,  
   the ISM is mainly heated by starlight and supplemented by the diffuse X-rays.  
However, CR heating begins to dominate shortly after the onset of violent star-formation. 
At the time the steady state is reached (by around 10~Myr), 
   the CR heating rates are $10^{-25}$, $10^{-24}$  and $10^{-23}$ erg cm$^{-3}$ s$^{-1}$~\footnote{
      These rates are substantially lower than those presented in~\cite{Owen2017}, 
      in which the heating effect was greatly overestimated  
      as the attenuation of CRs across the galaxy was not properly taken into account and the CR injection rate was overstated.}
   for ${\cal R}_{\rm SN} =  0.1$, 1 and $10~{\rm yr}^{-1}$ 
   respectively
  i.e. higher than the rates resulting from heating by starlight and diffuse X-rays. In the most intensely star-forming model, the final CR heating rate attained is slightly lower than might be expected from a direct scaling with $\mathcal{R}_{\rm SN}$. This is due to the comparatively elevated inverse-Compton loss rates of the secondary CR electrons and resulting lower heating efficiency, which arises from the higher intensity of the stellar radiation field in this case.
Note that some caution is required when comparing the CR and starlight heating with the X-ray heating.  
Fig.~\ref{fig:heating_sfr} shows 
   the X-ray heating corresponding to the case where the total X-ray power is $8.3\times 10^{38}~{\rm erg\;\!s}^{-1}$.  
The normalisation of the profile for diffuse X-ray heating 
    is scaled with the total power of the X-rays.  
Thus, the conclusion above should be modified if a higher value for the total X-ray power is adopted. 
For instance, if the X-ray power is around $8.3 \times10^{41}~{\rm erg\;\!s}^{-1}$, 
   X-ray heating will exceed starlight heating in the central 1~kpc of the galaxy.     
If the X-ray power reaches around $8.3 \times 10^{42}~{\rm erg\;\!s}^{-1}$, then  
   X-ray heating will become comparable 
   to CR heating for a SN event rate ${\cal R}_{\rm SN} = 0.1\;\!{\rm yr}^{-1}$.  We note that these radiative heating rates do not account for scattering or reprocessing in dust, which is unlikely to play as important a role in high-redshift starbursts than their low-redshift counterparts. See section~\ref{sec:dust} for further details.

\subsubsection{Energetics}
\label{sec:energetics}

The total radiative stellar and X-ray powers are around $2.8\times 10^{44}~\text{erg}~\text{s}^{-1}$ and $8.3 \times 10^{38}~{\rm erg\;\!s}^{-1}$ respectively, while that going into the CR primaries is around $3.0 \times 10^{40}~{\rm erg\;\!s}^{-1}$, i.e. 4 orders of magnitude less than the stellar radiation. The corresponding total power deposited into the ISM in the model is calculated to be $1.9\times10^{38}~\text{erg}~\text{s}^{-1}$ and $2.6 \times 10^{37}~{\rm erg\;\!s}^{-1}$ for stellar radiation and X-rays respectively while, for contained CRs, the heating power to the ISM is $2.1 \times 10^{39}~{\rm erg\;\!s}^{-1}$, or $7.5 \times 10^{32}~{\rm erg\;\!s}^{-1}$ in the free-streaming case (i.e. consistent with the $\sim 10^6$ level of CR containment within the galaxy). All these quoted values are for the conservative case, i.e. for ${\cal R}_{\rm SN} = 0.1~\text{yr}^{-1}$ (or around ${\cal R}_{\rm SF} = 16~\text{M}_{\odot}~\text{yr}^{-1}$ by the transformation between the two quantities from section~\ref{sec:rad_heating}), but are found to scale directly with SN (or SF) rate. The radiative (X-ray) heating power is around 4 orders of magnitude larger than that of the free-streaming CRs. However, when the containment of the CRs is established, it consistently follows that the radiative heating effect becomes sub-dominant, with the CR heating increasing by a factor of around $10^6$, to 2 orders of magnitude above the X-ray power.

\subsection{Discussion}
\label{sec:remarks}

\subsubsection{Comparison with Nearby Galaxies }
\label{sec:comparison}

Multiplying the value of $E{\rm d}N/{\rm d}E{\rm d}V$ in the steady state CR profile 
  (i.e. that in Fig~\ref{fig:diffusion_evolution})  
  with the mean energy of CRs ${\bar E}$ with energies above 1~GeV ($\approx 2.5~{\rm GeV}$) 
  gives an estimate for the CR energy density $\epsilon_{\rm CR}$ in the model galaxies. 
We show the results for three cases with SN event rates ${\cal R}_{\rm SN} =  0.1$, 1 and $10~{\rm yr}^{-1}$ 
   in Table~\ref{tab:cr_energy_densities}.   
Although, at present, we cannot be certain of the level of CR diffusion or containment in high-redshift galaxies, 
  comparison with values from observations of nearby examples offers some useful insight. 
\citet{YoastHull2015} have derived the CR energy densities for nearby galaxies,    
  including for the starbursts M82, NGC 253, and Arp 220. 
The CR energy densities of these three starbursts along with the Milky Way \citep[data from][]{Ferriere2001}, 
  and M31 \citep[data from][]{Abdo2010}  
  are also included in Table~\ref{tab:cr_energy_densities} for comparison.
  
\begin{table}
\begin{center}
  \begin{tabular}{ | c | c | }
    \hline
        \textrm{Source} & \textrm{CR energy density~/~eV cm$^{-3}$} \\ \hline
    Milky Way & 1.4 \\ 
    M31 & 1.5 \\
    M82 & 550~[1] \\
    NGC 253 & 260-350 \\
    Arp 220 & 1100-5100~[2]  \\
     \textrm{Model}: 0.1~${\rm SN}\ {\rm yr}^{-1}$ & 350 \\
   \textrm{Model}: 1~${\rm SN}\ {\rm yr}^{-1}$ & 3500 \\
\textrm{Model}: 10~${\rm SN}\ {\rm yr}^{-1}$ & 35000 \\
    \hline
    \end{tabular}
\end{center}
      \caption[Energy density comparison for CRs]
      {Energy densities of CRs $\epsilon_{\rm CR}$ in the Milky Way, M31 
       and three starburst galaxies M82, NGC 253, and Arp 220. 
        Data was taken from~\citet{Ferriere2001} for the Milky Way and from \citet{Abdo2010} for M31. 
        The others are obtained by \citet{YoastHull2015}.   
        The results for the three models are 
           for the case with full magnetic containment of the CRs as calculated for the stated SN rates. 
       [1] Estimated from $\gamma$-ray observations. 
           If combined with radio data, the CR density is then about $430-620 {\rm eV~cm}^{-3}$.  
      [2] Taking the full range of values across different regions of Arp 220 as in the study by \citet{YoastHull2015}.}
      \label{tab:cr_energy_densities}
\end{table} 

The CR energy density $\epsilon_{\rm CR}$ obtained for the model with ${\cal R}_{\rm SN} = 0.1~{\rm yr}^{-1}$ 
   agrees well with the observed values of  M82 and NGC 253, 
   which have a SN rate of $\sim 0.1~{\rm yr}^{-1}$ \citep{Lenc2006AJ, Fenech2010MNRAS}.  
The CR energy density $\epsilon_{\rm CR}$ obtained for the model with ${\cal R}_{\rm SN} = 1~{\rm yr}^{-1}$ 
  is also consistent with the observed value of Arp~220 
  which has a SN rate of $\approx 4~{\rm yr}^{-1}$ \citep{Lonsdale2006ApJ}. 
Although the model with the highest SN rate, at $10~{\rm yr}^{-1}$, yields 
  a CR energy density considerably higher than the three nearby starburst galaxies,  
  such an active galaxy may be representative of some of the most violent starburst protogalaxies present in the very distant Universe.    
By comparison, the Milky Way and M31 are expected to have very low CR energy densities 
   due to a relatively low level of star-formation activity, corresponding to a low SN event rate of 
   $\sim 0.015~{\rm yr}^{-1}$ \citep[see][]{Adams2013ApJ, Diehl2006Nat, Hakobyan2011Ap, Dragicevich1999MNRAS}. 
With such a rate, a direct scaling with the model with ${\cal R}_{\rm SN} = 0.1~{\rm yr}^{-1}$ 
  would suggest $\epsilon_{\rm CR} \sim 53~ {\rm eV~cm}^{-3}$, 
  more than an order of magnitude higher than the measured value for the Milky Way, at $1.4~{\rm eV~cm}^{-3}$.
This discrepancy arises as a scaling is required to geometrically account for the difference in the sizes of the galaxies. 
We may resolve this issue as follows: 
Take the steady-state form of equation \ref{eq:diff2} 
  and integrate it (by the divergence theorem).  
From equation~\ref{eq:cr_scale},    
  we then obtain by dimensional analysis 
\begin{equation} 
  4\pi R^2 \left(\frac{D}{R}\right) n =  \kappa~ {\cal R}_{\rm SN}   \ , 
\end{equation}   
  where $\kappa$ is some variable specifying the efficiency of CR production in a SN event.
This gives a more rigorous scaling relation as
\begin{equation}  
   \epsilon_{\rm CR,2} = \epsilon_{\rm CR,1} 
     \left(\frac{{\cal R}_{\rm SN,2}}{{\cal R}_{\rm SN,1}}\right) 
     \left(\frac{D_1}{D_2}\right)   \left(\frac{R_1}{R_2}\right)  \left(\frac{\kappa_2}{\kappa_1}\right) 
      \left(\frac{{\bar E}_2}{{\bar E}_1}\right)   \ . 
      \label{eq:scaling_relation}
\end{equation} 
Assuming that diffusion coefficients $D_1 \approx D_2$, the CR conversion variables $k_1 \approx k_2$  
  and the mean CR energies ${\bar E}_1 \approx {\bar E}_2$, 
  we obtain $\epsilon_{\rm CR,2} \approx 1.5~ {\rm eV~cm}^{-3}$ 
    if we set $\epsilon_{\rm CR,1} = 350 ~ {\rm eV~cm}^{-3}$, 
      $R_1 = 1~{\rm kpc}$ and ${\cal R}_{\rm SN,1} = 0.1~{\rm yr}^{-1}$ for the reference galaxy   
      and $R_2 = 30~{\rm kpc}$ \citep[see][]{Xu02015ApJ} 
      and ${\cal R}_{\rm SN,2} = 0.015~{\rm yr}^{-1}$ for the Milky Way. 
The CR energy density derived from the model and the measured value of the Milky Way thus reconcile.

\subsubsection{Comment on the Locality of Cosmic Ray Heating}
\label{sec:heating} 

Generally, for CR particle transportation, diffusion arises both in momentum (energy) space  
  as well as in physical space. 
Diffusion of charged particles in momentum space is usually caused 
   by radiative or scattering processes.  
Hadrons are inefficient emitters, thus
  their radiative loss timescales are considerably longer than those of other relevant processes.      
In galactic ISM environments, 
  the timescale of hadronic CR diffusion through momentum space 
  is expected to be much greater than the timescale of spatial diffusion \citep{Ko2017} or advection.   
The energy loss of a stream of CR particles 
   is therefore dominated by particle attenuation, such as that resulting from hadronic interactions. In the centre-of-momentum frame, 
   the secondary particles resulting from a hadronic interaction have lower energies and 
   shorter mean-free-paths (particularly when accounting for magnetic scattering) than their parent particles and we have shown that few interactions are required to entirely convert a hadronic primary into secondary pions (which decay to electrons and neutrinos). We note that neutrinos can only interact weakly so the amount of energy they can deposit in the 
protogalactic ISM or the IGM is insignificant. Thus the thermalisation process is entirely governed by the electronic secondaries, principally through Coulomb interactions and so the electron thermalisation timescale is estimated by the Coulomb timescale. 

A 1 GeV CR primary proton produces secondary electrons of energy $E_{\rm e} \approx 40$ MeV. In a 10 cm$^{-3}$ ionised ISM, the thermalisation timescale of these electrons then follows as
\begin{equation}
\tau_{\rm th} \approx 0.39 ~\left(\frac{E_{\rm e}}{40~\rm MeV}\right) \left(\frac{n_{\rm p}}{10~{\rm cm}^{-3}} \right)^{-1} ~\text{Myr} \ .
\end{equation}
If the secondary electrons propagated as freely-streaming particles with velocities close to the speed of light, ${\rm c}$, this would correspond to a thermalisation length-scale of around 0.1~Mpc -- considerably larger than the protogalactic host of characteristic radius 1~kpc. However, like the CR primaries, the secondary electrons are charged and their propagation in the magnetic field of the host is also more appropriately modelled as a diffusion process. The diffusion coefficient takes the same form as that for protons which, as the electrons are still relativistic when at tens of MeV, remains in its mass-independent form-- see e.g.~\cite{Kulsrud2005book}. The thermalisation length, $\ell_{\rm th}$ can therefore be estimated as the diffusion length arising over the timescale $\tau_{\rm th}$, i.e.
\begin{align}
\ell_{\rm th} &\approx \sqrt{4~D~\tau_{\rm th}} \nonumber \\
&\approx 0.17~\left(\frac{E_{\rm e}}{40~\text{MeV}}\right)^{3/4}\left(\frac{n_{\rm p}}{10~{\rm cm}^{-3}}\right)^{-1/2} ~\text{kpc} \ .
\end{align}
This shows $\ell_{\rm th}$ to be around an order of magnitude smaller than the characteristic size of the protogalaxy and so demonstrates that the CR energy is generally deposited well within the host. As such, we argue that it is sufficient to model the energy deposition as local to the initial CR interaction. Fully propagating the thermalisation lengths of the CR secondaries would simply spread the heating through a volume which remains well within the host's ISM, and so would not substantially alter the results of our calculation. Indeed, we note that the thermalisation length-scale is comparable to the resolution of our simulation anyway. In a model with sufficient resolution and explicitly accounting for the thermalisation length of the secondaries, differences from our present results would only emerge when ISM density substructures are included in the calculation -- in particular if modelling the  thermalisation within dense clouds compared to the hot, low-density component of the ISM. However, such matters fall beyond the scope of this paper, and are instead left to follow-up work.

\begin{figure}
	\includegraphics[width=\columnwidth]{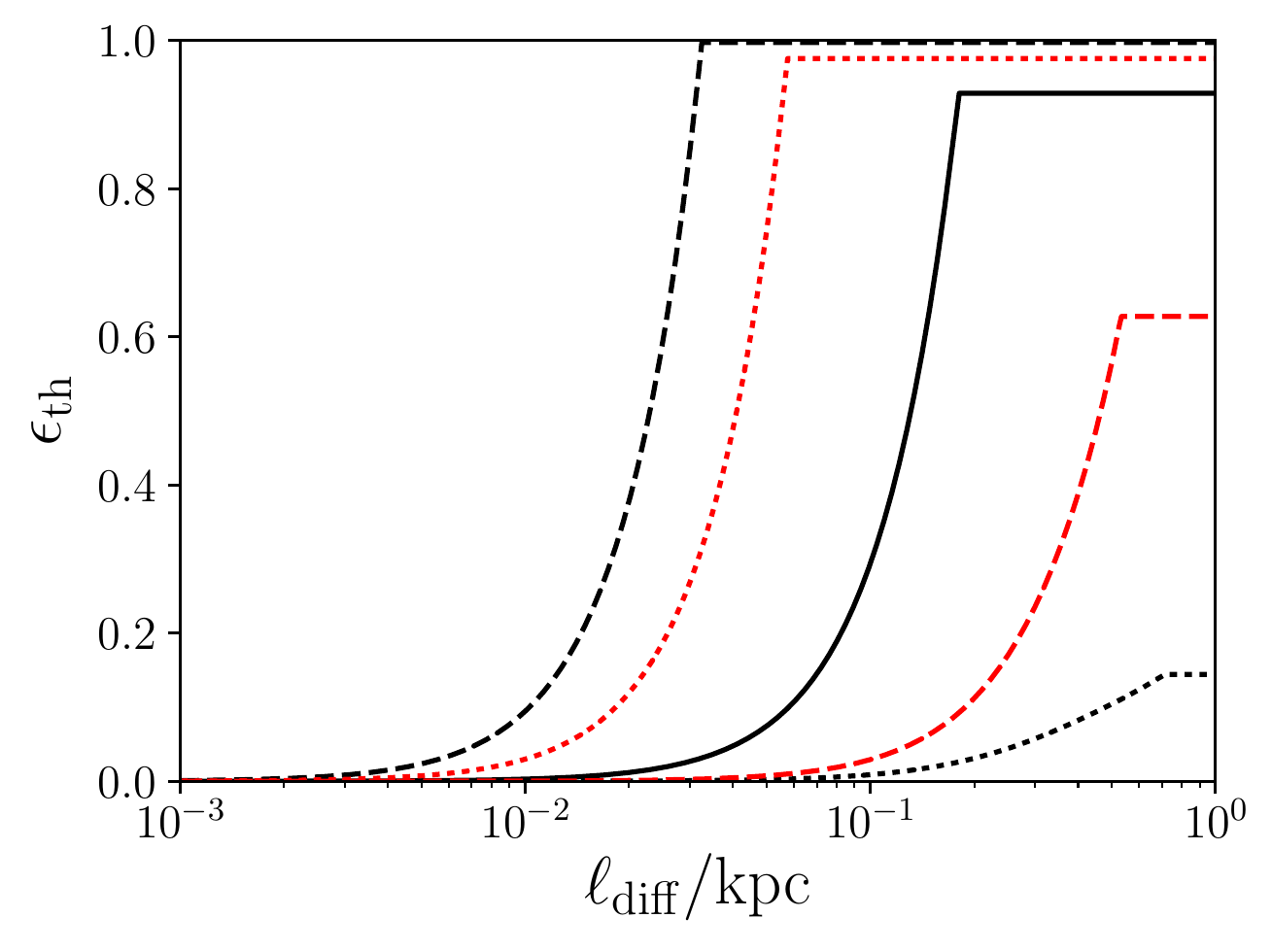}
\caption{Thermalisation fraction of a beam of CR secondary electrons along their diffusive propagations - this gives the fraction of the beam energy thermalised over a specified distance from a source. The reference (solid black) line gives the case for $E_{\rm e} = 40~\text{MeV}$ electrons traversing an ISM of density $n_{\rm p} = 10.0~\text{cm}^{-3}$. The dashed and dotted black lines then give the thermalisation fraction for $E_{\rm e} = 4~\text{MeV}$ and 400 MeV electrons respectively while keeping the density fixed at $n_{\rm p} = 10.0~\text{cm}^{-3}$. The red lines show the result for 40 MeV electrons in a medium of $n_{\rm p} = 1.0~\text{cm}^{-3}$ (dashed) and $100.0~\text{cm}^{-3}$ (dotted) respectively. Incomplete thermalisation of the beam (indicated by a plateau) shows the action of some electron energy loss by free-free and radiative processes, which do not directly contribute to the thermalisation process. The maximum thermalisation fractions achieved in each case (where the line plateaus) may also be read from Fig.~\ref{fig:timescale_v_energy} and~\ref{fig:timescale_v_density}. This uses stellar radiation fields consistent with $\mathcal{R}_{\rm SN} = 0.1~\text{yr}^{-1}$.}
\label{fig:thermalisation_length}
\end{figure}

In Fig.~\ref{fig:thermalisation_length}, we plot the thermalisation fraction of a beam of secondary CR electrons $\epsilon_{\rm th}$ over their diffusive propagation distance $\ell_{\rm diff}$ from their initial point of injection (i.e. the location of the initial CR primary interaction) for three energies and ISM densities. The reference case is given by a solid black line, appropriate for the protogalaxy model in this paper for CR electron secondaries of $E_{\rm e} = 40~\text{MeV}$ in an ISM of density $n_{\rm p} = 10~\text{cm}^{-3}$. The dashed and dotted black lines then give the thermalisation fraction over distance for $E_{\rm e} = 4~\text{MeV}$ and 400 MeV electrons respectively, while the dashed and dotted red lines show the result for 40 MeV electrons in a medium of $n_{\rm p} = 1.0~\text{cm}^{-3}$ and $100.0~\text{cm}^{-3}$ respectively. This demonstrates the important trends outlined earlier in section~\ref{sec:microphysics}: that higher energy CRs thermalise less efficiently (with $\epsilon_{\rm th}$ reaching less than 0.2 for the 400 MeV CR electrons) and do so over longer distances, and that CRs propagating though higher density environments thermalise more efficiently and more quickly - as indicated by the red dotted line where a 40 MeV CR in a medium just 10 times more dense than the reference model is able to thermalise very locally, within just a few tens of pc.

\subsubsection{Escape in a Cosmic Ray Wind}
\label{sec:leaking} 

In section~\ref{sec:results}, we found large increases in CR energy density could arise in the case of their magnetic containment in a protogalaxy ISM. If this is sufficient to yield a pressure gradient able to overcome the gravitational potential of the galaxy, then a galactic wind would erupt able to eject CRs into intergalactic space and drive an outflow. This could substantially reduce the number of CRs that could be harboured inside the host. It is therefore important to establish whether such a CR driven wind would emerge and, if so, whether it would have the potential to substantially reduce CR containment within the protogalaxy\footnote{While, in a more sophisticated model, this may also drive an increase in magnetic energy density, the essential matter is to determine whether the additional contained CR energy density is enough to drive a wind. Whether the energy density is truly harboured in CRs, or has been transferred in part to the magnetic field to an equipartition level by some mechanism is incidental because the coupling between the CRs and magnetic field means they would essentially act together as a single fluid exerting an effective pressure gradient against gravity. It may therefore be useful to think of the CR pressure/energy densities mentioned here as the combined CR and magnetic pressures and energy densities.}.

We model the wind as a conic section of a spherically symmetric outflow of a hydrogen fluid
   of density $\rho = m_{\rm p} n_{\rm p}$ and adiabatic index $\gamma_{\rm g} = 5/3$. 
The momentum equation governing the fluid flow is 
\begin{equation}
\frac{q}{r^2} \frac{{\rm d}v}{{\rm d}r}  = - \frac{{\rm d}P_{\rm c}}{{\rm d} r} - q \frac{G M(r)}{vr^4}
\label{eq:fluid2}
\end{equation}
where $q = \rho v r^2$ is the mass flow per unit time and $P_{\rm c}$ is the CR pressure. 
The adiabatic index of the CRs is $\gamma_{\rm c} = 4/3$.  
The energy equation is 
\begin{equation}
\frac{1}{r^2}\frac{\rm d}{{\rm d}r}\left[\frac{q v^2}{2}\right] = - q \frac{G M(r)}{r^4} + I \ ,
\label{eq:fluid_energy}
\end{equation}
and the energy evolution of the CRs is given by 
\begin{equation}
\frac{1}{r^2}\frac{\rm d}{{\rm d} r}\left[\left(\frac{\gamma_{\rm c}}{\gamma_{\rm c}-1}\right) P_{\rm c} v r^2\right] = -I
\label{eq:fluid3}
\end{equation}
\citep[][]{Salem2014MNRAS, Samui2010MNRAS, Ipavich1975ApJ, Drury1986MNRAS} where the interaction term $I$ introduced here quantifies the energy interchange between the CRs and the wind fluid. 
Without losing generality we obtain the Bernoulli equation 
  by combining equations~\ref{eq:fluid_energy} and~\ref{eq:fluid3}
\begin{equation}
\frac{q v^2}{2} + \left(\frac{\gamma_{\rm c}}{\gamma_{\rm c}-1}\right) P_{\rm c} v r^2 - q\frac{G M_{\rm gal}}{r} = C_{\rm e} \ ,
\label{eq:bernoulli}
\end{equation}
\citep[see][]{Samui2010MNRAS, Breitschwerdt1991A&A} 
  with $M_{\rm gal}$ introduced as the total galaxy mass, i.e. the convergent value of $M(r)$,
  and $C_{\rm e}$ being a constant determined by the boundary condition.     
The CR pressure relates to the CR energy density as $P_{\rm c} = (\gamma_{\rm c}-1)\epsilon_{\rm CR}$.  
It follows that\footnote{Note that equation~\ref{eq:dvdr} is identical to the expression given in \citet{Samui2010MNRAS}.}
\begin{equation}
\frac{{\rm d}v}{{\rm d}r} = \frac{2v}{r}\frac{\left[1-\frac{G M(r)}{2 r c_s^2}\right]}{\left[\frac{v^2}{c_s^2}-1\right]} \ , 
\label{eq:dvdr}
\end{equation}
   which enables the velocity of a cold, CR driven wind against gravity to be found. 
In the equation above, $c_s$ is the effective sound speed. For a CR dominated system, 
\begin{equation}
c_s^2 \approx \frac{\gamma_{\rm c} P_{\rm c}}{\rho} \approx \frac{\gamma_{\rm c}\epsilon_{\rm CR}}{3\rho} \ .
\end{equation}
The wind velocity is asymptotic to a terminal speed as $r\rightarrow \infty$. 
Such a wind is supersonic past its critical point, and  the sound speed must then be faster than the escape velocity $v_{\rm esc}$~\citep[the topology of the different types of solution are explored in, e.g.][]{Recchia2016MNRAS}. 
In equation~\ref{eq:bernoulli} (the Bernoulli equation), the constant $C_{\rm e}$ can be evaluated at a reference point, 
  e.g. the critical radius where the outflow becomes supersonic, and at $r\rightarrow\infty$~\citep{Breitschwerdt1991A&A}. 
 Assuming that the gravitational potential and the CR pressure as $r\rightarrow \infty$ are both substantially less than that at the critical point, 
  equation~\ref{eq:bernoulli} can be reduced to allow the asymptotic cold CR-driven wind velocity (hereafter the `terminal velocity') to be estimated as
\begin{equation}
v_{\rm CR, \infty}^2 \approx \frac{2 c_s^2}{(1-\gamma_{\rm c})}-v_{\rm esc}^2
\label{eq:vinfinity_val}
\end{equation}
\citep{Breitschwerdt1991A&A} where $v_{\rm esc}^2 = G M_{\rm gal}/r$.  We take the wind terminal velocity, $v_{\rm CR, \infty}$, as an upper limit for the rate at which CRs can leak out of the protogalaxy. From this, the CR escape timescale associated with any resulting wind, should one arise, can be estimated for the different CR energy densities and compared to the diffusion timescale for a given model. If the diffusion timescale is substantially shorter, then the CRs propagate predominately by diffusion and are effectively contained in the galaxy. If the wind escape timescale is shorter, then the CRs are able to effectively leak out of the host, and the containment picture no longer holds. 
For interim cases, where the timescales are comparable, both processes are important and the timescale ratios offer an appropriate scaling factor 
by which the CR containment level should be reduced to account for a non-negligible escaping fraction of particles, 
i.e. $f_{\rm esc} \approx \tau_{\rm esc}^{-1}/(\tau_{\rm esc}^{-1}+\tau_{\rm diff}^{-1})$, 
with $\tau_{\rm esc} \approx \ell_{\rm gal}/v_{\rm CR, \infty}$ and $\tau_{\rm diff} \approx \ell_{\rm gal}^2/4 D$.

Table~\ref{tab:table2}, shows the terminal velocity $v_{\rm CR, \infty}$ of cold CR-driven winds in various situations
   using the prescription described above. 
Note that there are cases where the CR driven outflow condition cannot be satisfied. 
Of the models, only the most violently star-forming galaxy with the highest SN rate of 10$~\text{yr}^{-1}$ 
  is able to drive a wind to enable the leaking of CRs. 
In this case, the escape fraction of the CRs does not exceed 50\%. 
Our earlier results (in section~\ref{sec:results}) omitting CR leaking are therefore not substantially different to those taking full account of CR leaking.  
However, based on the level of CR heating in the system investigated in section~\ref{sec:results} (whether or not accounting for CR escape), 
  we may conclude that CRs could be advected by a thermally driven outflow (see section~\ref{sec:advection}) 
  shortly after the onset of starburst activity. 
 The advection of CRs by such a wind and the consequences of this process will be discussed in \citealt{Owen2018_prep}.

\begin{table}
\begin{center}
  \begin{tabular}{ | c | c | c | }
    \hline
        \textrm{Source} & \textrm{$v_{\rm CR, \infty}$~/~km s$^{-1}$} & $f_{\rm esc}~/~\%$ \\ \hline
    Milky Way [1] & -- & -- \\ 
    M31 [2] & -- & -- \\
    M82 [3] & 700 & 92.7 \\
    NGC 253 [4] & 570 - 670 & 95.0 - 95.7 \\
    Arp 220 [5] & < 130 & < 33.4  \\
     \textrm{Model}: 0.1~${\rm SN}\ {\rm yr}^{-1}$ & -- & -- \\
   \textrm{Model}: 1~${\rm SN}\ {\rm yr}^{-1}$ & -- & -- \\
\textrm{Model}: 10~${\rm SN}\ {\rm yr}^{-1}$ & 390 & 49.8 \\
    \hline
    \end{tabular}
\end{center}
      \caption[Cold CR driven outflow wind capability of systems.]
      {Table to indicate the ability of systems to host a cold CR-driven outflow wind, choosing parameter values to favour the development of a wind. $v_{\rm CR, \infty}$ is the terminal velocity of a wind, should one be able to form, calculated using equation~\ref{eq:vinfinity_val} with the CR energy densities given in Table~\ref{tab:cr_energy_densities}. $f_{\rm esc}$ gives the fraction of CRs expected to escape from the galaxy due to the action of the wind, estimated from the escape timescale compared to the diffusion timescale. Systems which are gravitationally bound when considering their mass against the energy density of CRs cannot develop a cold CR driven outflow wind and no value is shown, although other driving mechanisms of outflows are still possible in which CRs may be advected (see section~\ref{sec:advection}) - e.g. thermal or line driven. Galactic size and total mass data used to calculate $v_{\rm CR, \infty}$ for [1] Milky Way size~\citep[][]{Xu02015ApJ}; total mass~\citep[e.g.][]{McMillan2017MNRAS, Kafle2014ApJ} [2] M31 size~\citep[][]{Chapman2006ApJ}; total mass~\citep[][]{Kafle2018MNRAS} [3] M82 size~\citep[see][for discussion]{Davidge2008AJ}; total mass~\citep[e.g.][]{Greco2012ApJ, Greve2002A&A} [4] NGC 253 size and total mass~\citep[][]{Bailin2011ApJ} [5] Arp 220 size and total mass~\citep[e.g.][]{Scoville1991ApJ}.}
      \label{tab:table2}
\end{table} 

\subsubsection{Dilution by Advective Transport}
\label{sec:advection} 

As per the empirical scaling outlined in section~\ref{sec:CR_diff}, the value of the diffusion coefficient is 
about $3\times 10^{28}$ cm$^2$ s$^{-1}$  for a 1 GeV CR when the magnetic field has reached saturation. 
With the length-scale of the system $\ell \approx 1$~kpc,  we can estimate the CR diffusion timescale as
\begin{equation}
    \tau_{\rm diff} \approx \frac{\ell^2}{4\;\!D} \approx 2.5~\text{Myr} \ . 
\end{equation}
Protogalaxies with strong star-formation activities are expected to exhibit wind outflows. 
These outflows carry CRs trapped by magnetic fields in the wind away
  from the host galaxy, thus reducing the amount of CR energy available for heating the ISM. 
The winds and outflows in nearby starburst galaxies 
  generally have bulk outflow velocities $v$ of around 10$^2$ to 10$^3$ km s$^{-1}$ 
  \citep[see][]{Chevalier1985Nat, Seaquist1985ApJ} 
  and the corresponding advection timescale of CR transport is 
\begin{equation}
   \tau_{\rm adv} \approx \frac{\ell}{v} \approx (1-10)~\text{Myr} \ .
\end{equation}

\begin{figure}
	\includegraphics[width=\columnwidth]{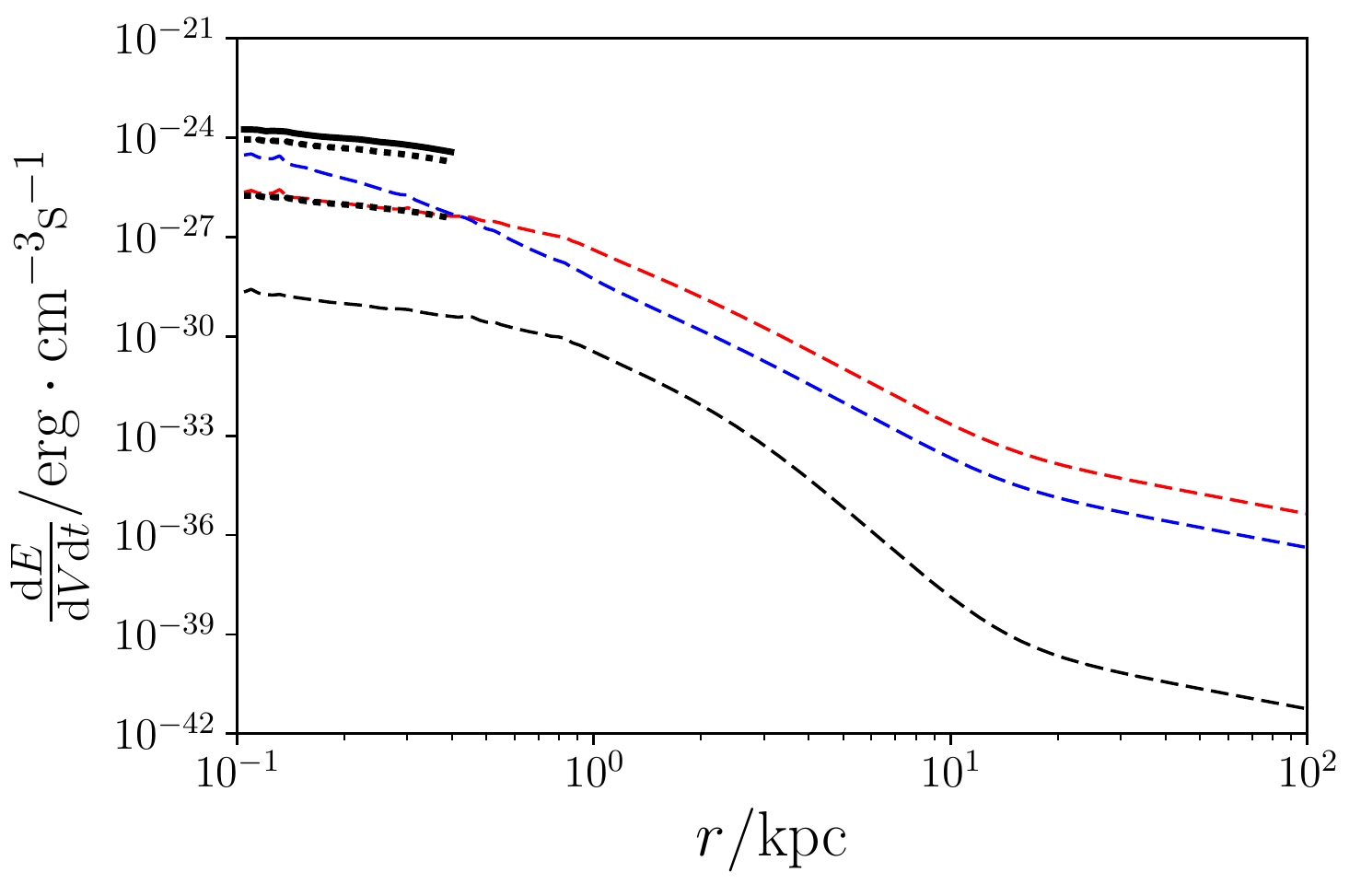}
\caption{
 Steady-state CR heating profiles for the case with 10, 50 and 99\% of the particles are removed by advection 
      (solid line, upper dotted line and lower dotted line in black respectively).    
   Heating profiles corresponding to starlight, X-rays and free-streaming CRs before magnetic containment occurs 
     are also shown (dashed lines in blue, red and black respectively) for comparison.  
   The other parameters are the same as those of the corresponding cases shown in Figure~\ref{fig:heating_sfr}.  All lines are calculated for ${\cal R}_{\rm SN} = 1.0~\text{yr}^{-1}$, corresponding to a star-formation rate of around $160~{\rm M}_\odot {\rm yr}^{-1}$ by the relation ${\cal R}_{\rm SF} \approx 160~{\rm M}_\odot {\rm yr}^{-1} ( {\cal R}_{\rm SN}/{\rm yr}^{-1})$ (see section~\ref{sec:mag_field_devel}).}
\label{fig:heating_advection}
\end{figure}

Given that the timescales of advection and diffusion are similar, 
  these processes must be comparably important in governing CR propagation 
  (and hence the energy budget) in ISM heating.  
The results shown in Fig.~\ref{fig:heating_sfr} and Table~\ref{tab:cr_energy_densities} 
  are obtained without the inclusion of advective transport of CRs. 
If a substantial fraction of CRs are carried away in advection by galactic outflows, 
  the CR heating of ISM will be reduced accordingly.  
Large-scale wind outflows in nearby starburst galaxies generally have a wide-angle cone morphology, 
   with the star-formation in the galactic disk below the cone providing the power. 
While advection is expected to dominate the CR transport process in the wind, 
  one would expect the heating of the ISM to be caused by the CRs diffusing across the galactic disk.  
Thus, to have a rough estimate of how much advective transport 
  is required to quench CR heating, 
  we may consider a simple prescription of removing a specific fraction of the CRs produced in the SN events, 
  attributing their loss to advective transport out of the environment. 
In Fig.~\ref{fig:heating_advection} we show the steady-state heating rate 
  for 10, 50 and 99\% of the CRs being removed (presumably) by the advective process 
  (at SN event rate of ${\cal R}_{\rm SN} = 1.0~{\rm yr}^{-1}$).   
We conclude that even when $99\%$ of the CRs are advected out of the system, 
   CR heating (which remains at least comparable with starlight and X-ray heating) 
   cannot be neglected when modelling the thermodynamics of the ISM in protogalaxies.    

\section{Discussions and Conclusions}
\label{sec:conclusions}

\subsection{Astrophysical Implications}  

In this study we have demonstrated that 
   CR heating power can exceed that due to conventional radiative heating 
   provided by starlight and diffuse X-rays
   for conditions appropriate to protogalaxies exhibiting strong star-formation activity. 
   This enhanced heating by CRs yields hotter proto-stellar gas,  
   while CR scattering and interactions 
   will induce additional turbulence throughout the ISM of the host 
   \citep[see, e.g.][]{Niemiec2008ApJ, Bykov2011, Rogachevskii2012ApJ}.  
The resulting additional (thermal and turbulent) pressure support will lead to increased Jeans masses in star-forming regions 
   \citep[see e.g.][]{Papadopoulos2011MNRAS, Papadopoulos2013}, 
   with proto-stars pushed to higher initial masses 
   \citep[see discussions in][]{Kuwabara2006ApJ, Ko2009ApJ, Hanawa2015ApJ, Hanawa2015PIAU, Kuwabara2015ApJ}. 
There are also suggestions that 
   severe CR heating can quench star-formation entirely in host environments, 
   perhaps even evaporating star-forming regions altogether in the most extreme cases 
   \citep{Guo2008MNRAS, Gabor2011MNRAS}.
Indeed, CRs may heat proto-stellar clumps throughout the host galaxy more effectively than radiation 
  due to higher levels of energy deposition at the point of interaction 
  and the possibility that they may be guided into star-forming regions 
  by the threading magnetic fields 
  \citep[see e.g.][]{Jokipii1966ApJ, Ferland2009}.  
In comparison with sub-keV X-rays and ultraviolet radiation, CRs are also less prone to shielding by the neutral material present in stellar nurseries and collapsing clouds and cores.
  
Increased containment of CRs within the host galaxy would be expected 
  to increase the rate of $\rm pp$-interactions driving the emission of $\gamma$-rays 
  \citep{Abdo2010, YoastHull2015, Wang2014, Wang2018MNRAS} 
  which is expected to be coupled with the CR heating effect 
  (see section~\ref{sec:interactions} for details). 
This may also be the case for synchrotron emission 
  following an enhanced number density of high energy CR secondary electrons 
  injected into dense regions of the environment by the primary CR proton interactions. 
These secondaries may then interact with the strengthening galactic magnetic field 
  \cite[see, e.g.][]{Schleicher2013A&A} and lose energy by synchrotron emission. 
At high redshift, the scattering of these energetic CR electrons off CMB photons 
  is also likely to drive a level of X-ray emission. 
The resulting X-ray illumination in high redshift starbursts is first introduced in 
   \citet{Schober2015MNRAS}, see also~\citet{Schleicher2013A&A}.

On larger scales in the environment outside the host galaxy, 
  the CR containment and the subsequent release of the CR energy  
  may drive large-scale galactic outflows and fountains 
    \citep{Shapiro1976ApJ, Bregman1980, Heesen2016MNRAS}. 
These outflows, in turn, may act as vehicles by which CRs are delivered into the circumgalactic medium 
  and beyond into intergalactic space \citep{Heesen2016MNRAS}. This process may be 
  a possible candidate for the pre-heating of the IGM in advance of cosmic reionisation 
  \citep{Sazonov2015MNRAS, Leite2017MNRAS}  
  at least in the proximity to a conglomeration of galaxies exhibiting violent star-forming activity.  
Additionally, the radiative emission from a host galaxy due to the contained CRs 
  - in particular the X-ray emission resulting from inverse-Compton scattering of electrons~\citep{Schober2015MNRAS} 
  - could begin to drive heating in the ISM of nearby neighbouring galaxies, 
  possibly impacting on their own internal processes via X-ray heating.

\subsection{Remarks}

Observations, via Zeeman effect measurements
   \citep[see e.g.][]{Bergin2007, Koch2015, Ching2017ApJ} 
   and dust emission polarimetry~\citep[e.g. see ][]{Planck2016bA&A}  
  have revealed complex magnetic structures in the Galactic ISM. 
The alignment and the spatial structure of magnetic fields in and around dense and/or star-forming regions 
  have shown great diversity 
  and also vary over different length-scales in the same system 
  \citep[see][]{Rao2009ApJ, Alves2011ApJ, Chen2012PIAU, Wang2015, Hull2017ApJ}. 
In the densest parts of some molecular clouds, 
  where stars are spawned,    
 the local magnetic field can reach levels of $\sim 1\;\!{\rm mG}$.  
  \citep{Crutcher2010ApJ}.   
Such field strengths are strong enough 
  to influence the propagation of CRs with energies even above a GeV, 
  which will have gyration radii $\sim 10^{11}{\rm cm}$  
(note that lower-energy MeV CRs will have smaller-scale gyration radii).  
The interaction of CRs with strong localised magnetic fields 
  therefore needs to be modelled more thoroughly  
  so as to determine their effects on CR scattering and propagation,   
  and how the energy-selected CR containment
  by such localised magnetic fields can impact on both the local and global 
  thermodynamics of the ISM of protogalactic environments.  

We have also examined the relative effect of starlight and X-ray heating.   
While intended as a comparison to the CR heating, 
   there is an intriguing interplay between the different processes 
   in addition to their relative importance in the thermodynamic context.  
For instance, the Compton scattering between the starlight photons 
  with the high-energy electrons which are produced by the interactions of the CR protons 
  may increase the level of X-rays inferred by stellar source counts, e.g. for SN remnants, X-ray binaries or colliding winds in stars. 
This effect may be particularly prevalent for violent star-forming protogalaxies at high redshifts 
  and may considerably enhance the `indirect' CR heating impact 
  on the interior and surrounding environment of the host.

We have employed a simple parametric model 
   as a demonstration of the importance of CR heating via hadronic processes in a protogalaxy. 
One of the crucial parameters in the model is the SN event rate, 
  which determines the total power available for the CR heating process, 
  be they confined or free-streaming. 
A uniform SN event rate is adopted in this study.  
The SN rate and magnetic fields are expected to co-evolve together 
  if galactic scale magnetism in the protogalaxy is seeded by SN events.    
In this coevolution scenario, 
  the containment of CRs in the initial and transient stage would be significantly different. 
However, one would not expect a substantial qualitative difference 
  between including and excluding such effect once the magnetic field evolution has saturated. 
Although we have assumed a uniform SN event rate, 
   the results obtained as such are nevertheless reasonable estimates.  

\subsection{Conclusions}

In this study we have shown that the containment of CRs 
   by a developing protogalactic magnetic field can arise within 10 Myr 
   if employing a SN turbulent-dynamo driven model as outlined in \citet{Schober2013A&A}. 
 Before this time, there is a window during which CRs may freely stream out of their host 
    and into their surrounding environment. 
 In the case of a protogalaxy actively forming stars 
   at a rate of around $\sim 160~{\rm M}_{\odot} {\rm yr}^{-1}$ 
   (leading to SN event rate of $\sim 1.0~{\rm yr}^{-1}$), 
   the calculated containment amplifies the expected ISM heating effect 
   due to the CRs from a level of around $10^{-29}$ ~erg ~cm$^{-3}$ s$^{-1}$ 
   (which is lower than the conventional radiative heating mechanisms by stellar emission and X-rays) 
   to around $10^{-24} ~{\rm erg~cm}^{-3}{\rm s}^{-1}$.  
Extrapolated from this level of CR heating within the central region of the protogalaxy, 
  an extra-galactic heating rate of $\sim 10^{-37}~{\rm erg~cm}^{-3}{\rm s}^{-1}$ 
  at around 10~kpc from the the protogalaxy, 
  and $\sim 10^{-41}~{\rm erg~cm}^{-3}{\rm s}^{-1}$ at 100~kpc from the protogalaxy is also expected.  
  
The level of CR heating in the protogalactic environment obtained here 
  also implies that CR heating in the IGM in the vicinity may also be non-negligible. 
This opens up many new questions about the impacts such an effect may have, 
  and the degree to which it could be maintained.

\section*{Acknowledgements}

We thank Prof Jim Hinton (MPIK), Dr Daisuke Kawata (UCL-MSSL), 
   Prof Shih-Ping Lai (NTHU), Dr Ignacio Ferreras (UCL-MSSL) for helpful discussion.  
We also thank Prof Luke Drury (DIAS) and Prof Chung-Ming Ko (NCU) 
  for highlighting the the competition between CR advection, CR diffusion and radiative energy losses 
  in galactic environments  
  and how CR attenuation along their propagations is crucial in determining the total energy budget of the system. 
Additionally ERO thanks Prof Jim Hinton and Prof Dr Werner Hofmann (MPIK) 
  for initiating his interest in cosmic ray astrophysics 
  and for encouragement in the early stages of this work.  
We are also grateful for the critical but helpful comments from the referee, 
   which led us to substantially refine our calculations 
   so that the system is now properly and self-consistently modelled. 
ERO is supported by a UK Science and Technology Facilities Council Studentship 
  and by the International Exchange Scholarship of NTHU (National Tsing Hua University), 
   kindly hosted by Prof Shih-Ping Lai. 
PS was supported by the State of Mauritius Postgraduate Scholarship while at University College London. 
This research has made use of NASA's Astrophysics Data Systems.




\bibliographystyle{mnras}
\bibliography{references} 



\appendix

\section{Energy Density of Spatially Extended Sources}
\label{sec:appendixa}

To extend the model radiation field from a single point source to an ensemble of $N$ sources, we introduce a distribution function $f(r)$ for the sources in a sphere. The effective distance between the nearest source and some observer located within the spherical distribution is $\tilde{r}$. In an axisymmetric system, the minimum distance to a source from a randomly positioned observer is $\tilde{r}$ while the maximum distance is the radius of the extended source, $r_{\rm sph}$, with the average energy density due to the radiation from the entire distribution of sources in the ensemble as
\begin{align}
U &= \frac{L}{4 \pi \rm{c}}\int_{\tilde{r}}^{r_{\rm sph}} {\rm d}r 4\pi r^2 \frac{f(r)}{r^2} \nonumber \\
&= \frac{L}{\rm{c}}\int_{\tilde{r}}^{r_{\rm sph}}{\rm d}r f(r) \ ,
\end{align}
where the distribution function $f(r)$ for a uniform distribution of sources is
\begin{equation}
f(r) = \frac{3}{4\pi}{N}\left({r_{\rm gal}^3-\tilde{r}^3}\right) \ .
\end{equation}
This yields
\begin{align}
\label{eq:energydens}
U &=\frac{3}{4\pi}\frac{L}{\rm{c}}N\left(\frac{r_{\rm sph}-\tilde{r}}{r_{\rm sph}^3 - \tilde{r}^3}\right) \nonumber \\
&=\frac{3}{4\pi}{N}{r_{\rm sph}^2}\left[1+\left(\frac{\tilde{r}}{r_{\rm sph}}\right) + \left(\frac{\tilde{r}}{r_{\rm sph}}\right)^2\right]^{-1} \ .
\end{align}
The number of sources within the ensemble can be calculated by the ratio of volume taken up by the source 
   (based on the nearest distance from the observer) to the size of the ensemble,
\begin{equation}
N = \frac{4\pi r_{\rm sph}^3}{4 \pi \tilde{r}^3} = \left(\frac{r_{\rm sph}}{\tilde{r}}\right)^3 \ ,
\end{equation}
which implies
\begin{equation}
\label{eq:ratio_rs}
\frac{\tilde{r}}{r_{\rm gal}} = \left(\frac{1}{N}\right)^{1/3} \ .
\end{equation}
Then, substitution of~\ref{eq:ratio_rs} into equation~\ref{eq:energydens} gives
\begin{equation}
\label{eq:radresult}
U = \frac{3}{4\pi}\frac{L}{\rm{c}} \frac{N}{r_{\rm sph}^2}\left[1+\left(\frac{1}{N}\right)^{1/3}+\left(\frac{1}{N}\right)^{2/3}\right]^{-1} \ .
\end{equation}
In the limit of $N\rightarrow 1$, the usual inverse-square law result for radiation can be recovered, i.e. $U=L/4\pi {\rm{c}} r_{\rm sph}^2$. For large $N$, the result becomes
\begin{equation}
\label{eq:raddensity}
U \rightarrow \frac{LN}{4 \pi \rm{c}}\left(\frac{3}{r_{\rm sph}^2}\right) \ .
\end{equation}
Comparison with an MC approach confirms the analytical result above for the energy density at randomly selected points in a spherical distribution of $N$ sources.
Division of equation~\ref{eq:radresult} by the typical photon energy of $2.82 \;\! {k_{\rm B}} T_{\rm *} \approx 3 \;\! {k_{\rm B}} T_{\rm *}$ yields an estimate for the photon number density of the radiation field in the case of stars distributed in a galaxy as 
\begin{equation}
\label{eq:result_radfield}
n_{\rm ph} \approx \frac{3}{12\pi k_{\rm B} T_{*}}\frac{L}{\rm{c}} \frac{N}{r_{\rm gal}^2}\left[1+\left(\frac{1}{N}\right)^{1/3}+\left(\frac{1}{N}\right)^{2/3}\right]^{-1} \ .
\end{equation}

\section{Hadronic Absorption of Cosmic Rays with Branching}
\label{sec:appendixb}

The locality of energy deposition of a CR particle can be justified as follows. 
Suppose, in a hadronic interaction, a fraction $f$ of the energy in a baryon beam 
  is retained while the rest of the energy is deposited into the target medium. 
The effective path-length for energy deposition in the presence of branching is then 
\begin{eqnarray} 
  {\bar  r_{\rm p\pi}} & = &\frac{ \sum_{{\rm k} =1} k\;\!  r_{\rm p\pi} \;\! W_{\rm k}(f)}{\sum_{{\rm k} =1} W_{\rm k}(f)} \ , 
\end{eqnarray}  
 where $r_{\rm p\pi}$ is the interaction mean-free-path for the dominant interaction (i.e. ${\rm pp}$).    
Here, the energy weighting factor is $W_{\rm k} = f^{\rm k-1}(1-f)$,  
 and it is normalised, i.e.  
\begin{eqnarray}      
  {\sum_{{\rm k} =1} W_{\rm k}(f)} & = & 1 \ .  
\end{eqnarray}  
It follows that 
\begin{eqnarray} 
\label{eq:path_length_b3}
  {\bar  r_{\rm p\pi}} & = & r_{\rm p\pi}(1-f)+2\;\!r_{\rm p\pi} f(1-f) \nonumber \\
   & & + 3\;\!r_{\rm p\pi} f^2 (1-f)+ 4\;\!r_{\rm p\pi} f^3 (1-f) +  \cdots  \nonumber  \\ 
    & = &  r_{\rm p\pi} (1-f)\left( \frac{\rm d}{{\rm d}f} \sum_{\rm k=1} f^k \right)   \nonumber \\  
    & = & { r_{\rm p\pi}}\;\!{(1-f)^{-1}}  \ .  
\end{eqnarray} 
The elasticity (i.e. energy which is retained by hadrons after an interaction) of the ${\rm pp}$ interaction is around $f = 0.3$. 
By equation~\ref{eq:path_length_b3}, this gives the effective interaction mean free path as $10 ~r_{\rm p\pi}/7$.  
For containment of the charged CR particles within the galaxy, $r_{p\pi}/c \ll  t_{\rm con}$, the particle containment time scale. 
Hence, the approximation of total energy transfer to secondary pions at the first point of hadronic interaction is justified.  

\bsp	
\label{lastpage}
\end{document}